\documentclass[11pt,a4paper]{article}
\pdfoutput=1
\usepackage{jheppub}
\usepackage[utf8]{inputenc}
\usepackage{tabularx}
\usepackage{array}
\usepackage{graphics}
\usepackage{graphicx}
\usepackage{psfrag}
\usepackage{epsfig}
\usepackage{amsmath}
\usepackage{amssymb}
\usepackage[normalem]{ulem}
\usepackage{setspace}
\usepackage{rotating}
\usepackage{colortbl}
\usepackage{tabularx}
\usepackage{longtable}
\usepackage{slashed}
\makeatletter
\usepackage{textcomp}
\usepackage[usenames,dvipsnames]{xcolor}
\usepackage{relsize}

\setlongtables

\newcommand{\bea}{\begin{eqnarray}}
\newcommand{\eea}{\end{eqnarray}}
\newcommand{\beq}{\begin{equation}}
\newcommand{\eeq}{\end{equation}}
\newcommand{\ec}{\end{center}}
\newcommand{\bc}{\begin{center}}

\newcommand{\pdir}{p\kern -5.2pt\raise 0.2ex\hbox {/}}

\newcommand{\vdir}{v\kern -5.75pt\raise 0.15ex\hbox {/}}
\newcommand{\kdir}{k\kern -5.75pt\raise 0.15ex\hbox {/}}
\newcommand{\epsdir}{\epsilon\kern -5.0pt\raise 0.15ex\hbox {/}}
\newcommand{\bvdir}{\bar{v}\kern -5.75pt\raise 0.15ex\hbox {/}}
\newcommand{\Ddir}{D\kern -7.75pt\raise 0.20ex\hbox {/}}
\newcommand{\Adir}{A\kern -7.75pt\raise 0.20ex\hbox {/}}
\newcommand{\ldir}{l\kern -5.0pt\raise 0.2ex\hbox{/}}
\newcommand{\varepsdir}{\varepsilon\kern -5.5pt\raise 0.15ex\hbox{/}}

\newcommand{\cb}{{\cal B}}

\title{Seeking leptoquarks in IceCube}

\author[a]{Damir Be\v{c}irevi\'c,} 
\author[b]{Boris Panes,} 
\author[c,d]{Olcyr Sumensari} 
\author[b]{and Renata Zukanovich Funchal}

\affiliation[a]{Laboratoire de Physique Th\'eorique (B\^at.~210)\\
Universit\'e Paris-Sud and CNRS (UMR 8627), F-91405 Orsay-Cedex, France}
\affiliation[b]{Instituto de F\'isica, Universidade de S\~ao Paulo, 
 C.P. 66.318, 05315-970 S\~ao Paulo, Brazil}
\affiliation[c]{Dipartimento di Fisica e Astronomia `G. Galilei', Universit\`a di Padova, Italy}
\affiliation[d]{Istituto Nazionale Fisica Nucleare, Sezione di Padova, I-35131 Padova, Italy}

\abstract{
We investigate the sensitivity of IceCube(-Gen2) to a scalar leptoquark scenario with couplings only to heavy quark flavors which may be connected to solving discrepancies in $B$-meson semileptonic decays. We take into account, for the first time, the non-negligible neutrino-gluon cross section induced by leptoquarks, and we systematically account for indirect and direct constraints which have been overlooked in previous studies. We conclude that IceCube(-Gen2) can only probe the light LQ regime, already disfavored by the combination of flavor physics constraints, electroweak precision data and the direct searches at the LHC.

}

\arxivnumber{}

\begin{document}
\begin{flushright}
\begin{tabular}{l}
{\tt \footnotesize LPT-Orsay-18-42}\\
\end{tabular}
\end{flushright}

\maketitle

\section{Introduction}
\label{sec:0}
It is remarkable that the two types of constituents in the flavor
sector of the Standard Model (SM), quarks and leptons, although
independent, share a number of common features. They come in three
generations that conspire to exactly cancel the triangle anomalies of
the gauge interactions, they exhibit mixing between mass and flavor
eigenstates resulting in oscillations, and flavor changing
neutral interactions do not occur at tree-level.  These common features strongly suggest that quarks and leptons should be somehow interrelated. In fact, leptoquarks (LQs), bosons that can mediate quark-lepton transitions, appear in many extensions of the SM, in particular, in Grand Unified Theories~\cite{Georgi:1974sy,Pati:1974yy}, and composite Higgs models~\cite{Abbott:1981re,Schrempp:1984nj,Wudka:1985ef}.

From a phenomenological point of view, LQs are $SU(3)_c$ colored particles
that can come as $SU(2)_L$ singlets, doublets or
triplets~\cite{Buchmuller:1986zs,Dorsner:2016wpm}. Their masses, 
strength and flavor structure of their couplings to quarks and
leptons are all undetermined when the ultraviolet completion is
unknown.  Since LQs with masses of the order of the electroweak scale
are not disallowed by any fundamental reason, it is important to look
for them in all possible ways.  In fact, many collider experiments
have been searching for them in pair as well as in single production
modes.  From HERA to the LHC, experiments have been setting stringent
limits on their masses and couplings,
c.f.~\cite{Aktas:2005pr,Diaz:2017lit} and references therein. These
limits depend, unfortunately, unavoidably on the assumed flavor
structure, some of which are difficult to explore in
accelerators. Other bounds deduced from the direct searches have been
discussed in Ref.~\cite{Greljo:2017vvb}, see also references therein.

Besides the purely theoretical reasons that compel experimentalists to look for these particles 
it seems there might be an experimental one too, coming from $B$-physics. The excellent agreement between the SM
predictions and experimental data seems to be disrupted by flavor
physics. Recent data from the LHC, Belle and BaBar
suggest lepton flavor universality violation
(LFUV) in $B$-meson semileptonic decays. The LHCb experiment, for
instance, has reported about 2.6$\sigma$ discrepancy in the ratio of the
partial branching fractions of the loop induced transitions (denoted by $\mathcal{B}^\prime$), 
\begin{equation}
R_{K^{(\ast)}}^{\rm exp} = \frac{{\cal B}^\prime(B \to K^{(\ast)} \mu^+ \mu^-)}{{\cal B}^\prime(B \to  K^{(\ast)}  e^+ e^-)} \, ,
\label{eq:RK0}
\end{equation}
and the measured values~\cite{Aaij:2014ora,Aaij:2017vbb} appear to be smaller than predicted in the SM~\cite{Hiller:2003js,Bordone:2016gaq}. 
Another intriguing indication of LFUV comes from the tree-level ratio~\cite{Amhis:2016xyh}
\begin{equation}
R_D^{\rm exp} = \frac{{\cal B}(B \to D \tau \nu)}{{\cal B}(B \to D \ell \nu)} = 0.41 \pm 0.05\, ,
\label{eq:RD}
\end{equation}
where $\ell=e,\mu$, obtained by combining Babar and Belle results~\cite{Lees:2012xj,Huschle:2015rga,Sato:2016svk,Aaij:2015yra,Hirose:2016wfn} and 2.2$\sigma$
above $R_D^{\rm SM}= 0.300(8)$, the SM prediction~\cite{Amhis:2016xyh,Na:2015kha,Lattice:2015rga}. This feature was confirmed with $R_{D^\ast}^{\rm exp} = 0.304(15)$~\cite{Lees:2012xj,Huschle:2015rga,Aaij:2017deq}, which also appears about 3$\sigma$ larger than the SM prediction,  $R_{D^\ast}^{\rm
  SM}= 0.260(8)$~\cite{Fajfer:2012vx,Bigi:2017jbd}.  
 Another confirmation of this phenomenon was reported by LHCb who measured $R_{J/\psi}=\cb(B_c\to J/\psi\tau \nu)/ \cb(B_c\to J/\psi\mu \nu)=0.71\pm 0.25$~\cite{Aaij:2017tyk}, 
 also larger than the SM estimates although the SM value in this case is less reliable. Many proposed solutions to these anomalies discussed in the literature so far rely on existence of LQ states~\cite{Hiller:2014yaa,Ghosh:2014awa,Alonso:2015sja,Fajfer:2015ycq,Barbieri:2015yvd,Bauer:2015knc,Deppisch:2016qqd,Arnan:2016cpy,Becirevic:2016yqi,Cox:2016epl,Hiller:2016kry,Dorsner:2017ufx,Barbieri:2016las,Becirevic:2017jtw,Fajfer:2018bfj,Cai:2017wry,Aloni:2017ixa,Crivellin:2017dsk,Crivellin:2017zlb,Buttazzo:2017ixm,DiLuzio:2017vat,Calibbi:2017qbu,Bordone:2017bld,Barbieri:2017tuq,DiLuzio:2017fdq,Cline:2017aed,Blanke:2018sro,Assad:2017iib},  notice that similar solutions can be implemented by considering supersymmetric models including R-Parity violating squarks~\cite{Altmannshofer:2017poe}. The
couplings of LQs to heavy quarks can be as peculiar as to make them difficult for direct observations in colliders and it is therefore useful to try and look for them in other ways.

Here we study a possibility of using the IceCube neutrino telescope in the South
Pole as a complementary way to search for those states.  Many authors
have explored the potential of IceCube to test LQ models, in
particular, after the observation of an excess of high energy stating
events consistent with a flux of high-energy astrophysical neutrinos
from outside our galaxy~\cite{Aartsen:2013jdh}. In
Ref.~\cite{Anchordoqui:2006wc} electroweak singlet LQ that
couple to either second or third generation of quarks and/or leptons were
studied, emphasizing the use of the inelasticity distribution of
events to improve background rejection.  Refs.~\cite{Barger:2013pla},
\cite{Dutta:2015dka} and \cite{Mileo:2016zeo} evoke LQ
couplings to the first generation of quarks and leptons in order to try to
accommodate IceCube high-energy neutrino data. Similarly, in \cite{Dev:2016uxj} a supersymmetric model including 
R-parity violating squark resonances is considered to explain IceCube data. The limits that
IceCube data can impose on a specific low-scale quark-lepton
unification model with scalar LQs was studied in
\cite{Dey:2015eaa}. Scalar LQ doublets that couple the first
and third family of quarks to leptons, with a flavor structure that
can address the anomalies in 
$R_{K^{(\ast)}}$, pass other LHC constraints and give a sizable
contribution to IceCube neutrino data were considered in
\cite{Chauhan:2017ndd} and by a more elaborated analysis in
\cite{Dey:2017ede}. The LQ states in all of these works, with
exception of Ref.~\cite{Anchordoqui:2006wc}, couple to the first
generation of quarks. Couplings to the first generation suffer,
however, from severe experimental constraints as they have to comply
with limits from direct searches at colliders, atomic parity violation
experiments and flavor physics observables in the kaon/pion sectors. Moreover, all previous studies of the neutrino-nucleon cross-section at IceCube only consider the resonant $s$- and
$t$-channel interactions of high energy neutrinos with quarks (valence and/or sea) via a LQ exchange. Nonetheless, we should not forget that in order to compute neutrino-nucleon cross sections we must also include the neutrino interaction with gluons mediated by a LQ, particularly important in accessing LQs that only couple to heavy quarks. 
Those are, in fact, the kinds of LQs that can take part in solving the discrepancies between data and heavy meson semileptonic decays. 
In this paper, we include for the first time the neutrino-gluon cross-section in order to consistently study the sensitivity 
of IceCube to LQ scenarios. Furthermore, we systematically compare the future IceCube future sensitivity with the indirect constraints coming from flavor physics and $Z$-pole observables, as well as the direct limits from the LHC, which are very efficient in constraining the LQ couplings.

\

Our paper is organized as follows. In Sec.~\ref{sec:1} we describe a LQ model with a particular simple flavor
structure that can solve $R_{D^{(\ast)}}$ and give rise to new neutrino-nucleon interactions which can be probed by IceCube. 
Furthermore, we perform a complete study of the indirect bounds on LQ couplings stemming from flavor physics 
and $Z$-pole observables. In Sec.~\ref{sec:3} we review the High Energy Starting Events sample collected by IceCube after six years 
and we discuss the current status of the SM fit. We explain the procedure to simulate both the SM 
and LQ events at IceCube in order to compare with experimental data. In Sec.~\ref{sec:4} we present results of our analysis and 
discuss what we can learn about our model from six years of IceCube data as well as the prospects for what can be
expected in the future. We finally conclude in Sec.~\ref{sec:conc}. 
All LQ contributions to the neutrino-nucleon 
scattering cross section, including the new ones, are collected in Appendix~\ref{lqxsec}, while in Appendix~\ref{transport}
we describe the transport equation for Earth's attenuation. The decay rate expressions for $B\to D \ell \bar{\nu}$ are provided in Appendix~\ref{app:flavor}.

\section{\label{sec:1} A Simple Leptoquark Scenario to Explain $R_{D^{(\ast)}}$}

In this paper we focus on the so called $R_2$-model which involves an electroweak doublet of scalar LQs with hypercharge $Y=7/6$, 
described by the Yukawa Lagrangian, 
\begin{align}
\label{eq:slq1}
\begin{split}
\mathcal{L}_{\Delta^{(7/6)}} &= (g_R)_{ij}
\bar{Q}_i{\boldsymbol\Delta}^{(7/6)}\ell_{Rj}+ (g_L)_{ij} \bar{u}_{Ri}
    { \widetilde{\boldsymbol\Delta}}^{(7/6) \dagger} L_{j}+
    \mathrm{h.c.}\\[2.ex] 
    &=(V g_R)_{ij}\bar{u}_i P_R
    \ell_j\,\Delta^{(5/3)}+(g_R)_{ij}\bar{d}_i P_R
    \ell_j\,\Delta^{(2/3)} \\[1.3ex] &\quad + (g_L U)_{ij}\bar{u}_i
    P_L \nu_j\,\Delta^{(2/3)}-(g_L)_{ij}\bar{u}_i P_L
    \ell_j\,\Delta^{(5/3)} +\mathrm{h.c.},
\end{split}
\end{align}
where in the first line the Lagrangian is given in the flavor basis and the superscripts in the LQ fields refer to the hypercharge ($Y$), while in the other two lines the Lagrangian is given in the 
mass eigenstate basis in which the superscripts of LQs refer to the electric charge $Q=Y+T_3$, 
where $Y$ is the hypercharge and $T_3$ the third component of weak isospin. Note that ${\widetilde{\bold\Delta}}= i \tau_2 \bold\Delta^\ast$ is the conjugate of the doublet of mass degenerate LQs, while 
$V$ and $U$ stand for the Cabibbo-Kobayashi-Maskawa (CKM) and the
Pontecorvo-Maki-Nakagawa-Sakata (PMNS) matrices,
respectively. 
Furthermore, $P_{L,R}=(1\mp\gamma_5)/2$ are the chiral projectors, $Q_i =
[(V^\dagger u_L)_i\; d_{Li}]^T$ and $L_i = [(U\nu_L)_{i}\;
  \ell_{Li}]^T$ denote quark and lepton SU(2)$_L$
doublets, whereas $u_L$, $d_L$, $\ell_L$ and $\nu_L$ are the fermion mass eigenstates.

We opt for the minimalistic structure of the Yukawa coupling matrices $g_{L,R}$ and assume
\bea
\label{eq:YC}
g_{L} = \left( \begin{matrix}
  0 & 0 & 0\\
  0 & 0 & g_{L}^{c \tau}\\
  0 & 0 & 0
\end{matrix}\right), \qquad g_{R} = \left( \begin{matrix}
  0 & 0 & 0\\
  0 & 0 &0\\
  0 & 0 & g_{R}^{b \tau}
\end{matrix}\right), \qquad V g_{R} = \left( \begin{matrix}
  0 & 0 & V_{ub} g_{R}^{b \tau}\\ 0 & 0 & V_{cb} g_{R}^{b \tau}\\ 0 &
  0 & V_{tb} g_{R}^{b \tau}
\end{matrix}\right), 
\eea
that allows $\Delta^{(2/3)}$ to mediate a tree-level contribution to 
$R_{D^{(\ast)}}$ and generates a contribution to the neutrino-nucleon scattering in IceCube. 

\

\subsection{$R_2$ explanation of $R_{D^{(\ast)}}$}
\label{sec:RD}

\begin{figure}[t!]
\centering
\includegraphics[width=0.25\linewidth]{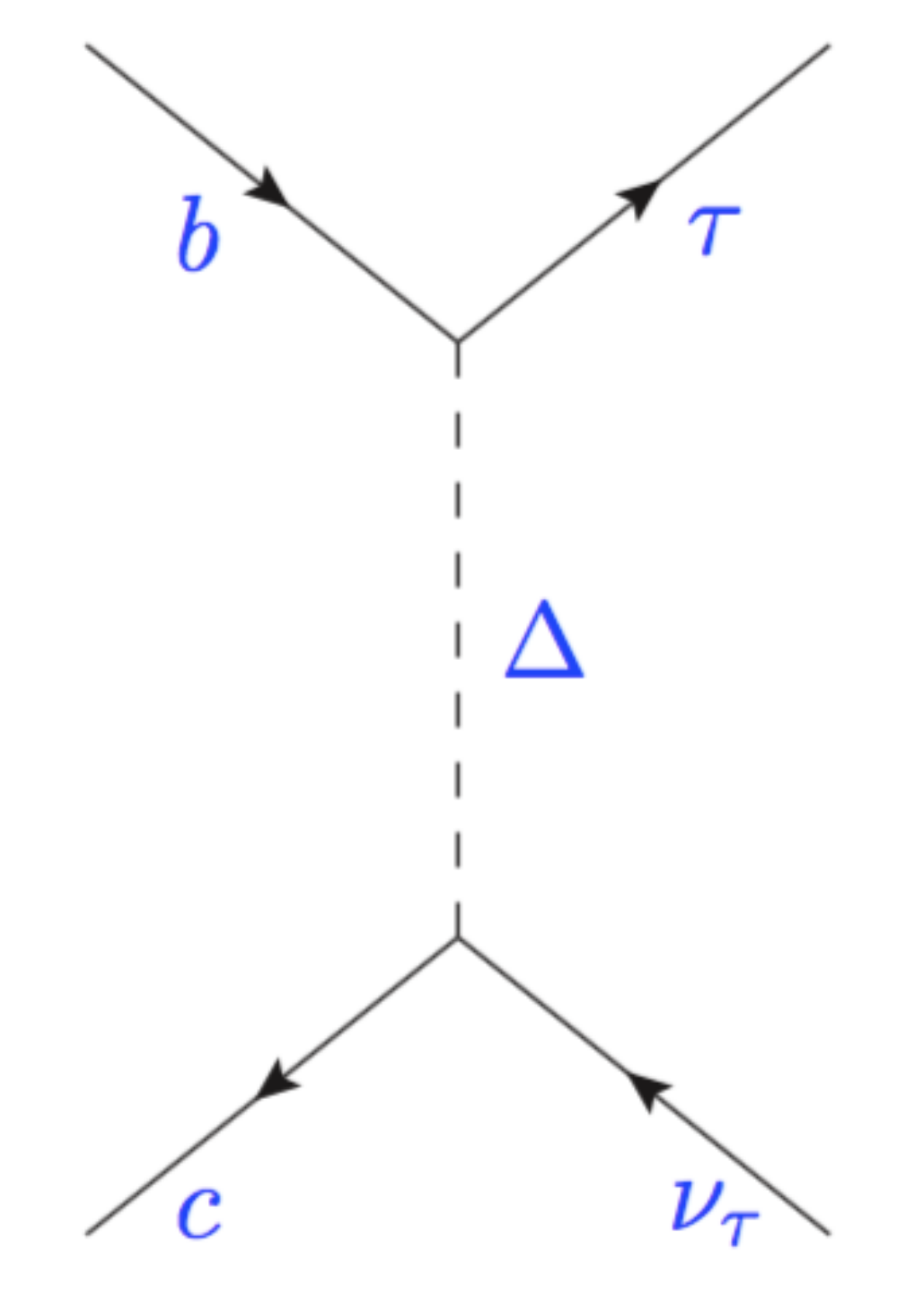}
\caption{\small \sl Contribution to $b\to c\tau\bar \nu_\tau$ arising from the LQ model discussed in this paper. $\Delta \equiv \Delta^{(2/3)}$ stands for the LQ.}
\label{FIG:semil_LQ}
\end{figure}

In order to confront the LQ contributions with the experimental data involving the $B \to D \tau \nu$ decay we consider the effective Hamiltonian
\begin{align}
\begin{split}
{\cal H_{\mathrm{eff}}} = \frac{4 \, G_F}{\sqrt{2}} V_{cb} \bigg{[} 
(\bar{\tau}_L \gamma^\mu \nu_L)(\bar{c}_L  \gamma_\mu b_L) 
&+ g_S(\mu)\, (\bar{\tau}_R \nu_L)(\bar{c}_R  b_L) \\[0.05em]
&+ g_T(\mu)\, (\bar{\tau}_R \sigma^{\mu \nu}\nu_L)(\bar{c}_R  \sigma_{\mu \nu} b_L) 
\bigg{]} + \mathrm{h.c.} \,,
\end{split}
\label{eq:hamiltonian}
\end{align}

\noindent where $g_{S,T}$ are the Wilson coefficients induced by the LQ state mediating the semileptonic decay via a tree-level contribution shown in Fig.~\ref{FIG:semil_LQ}. 
After integrating out $\Delta^{(2/3)}$, the expression for $g_{S,T}$, at the matching scale $\mu=m_\Delta$, reads:
\bea
g_S(\mu = m_\Delta) = 4 \, g_T(\mu = m_\Delta) = \frac{g_L^{c\tau}\, \left(g_R^{b\tau}\right)^{\ast}}{4 \sqrt{2} \, m_\Delta^2 \, G_F V_{cb}}\, ,
\label{eq:gsgt}
\eea
clearly enhanced by $V_{cb}^{-1}$. By virtue of the leading order QCD running from $\mu= m_\Delta \approx 1$~TeV down to $\mu=m_b$,  the above relation between $g_S$ and $g_T$ becomes
\bea
\label{eq:running}
g_S(\mu= m_b)  \approx 7.2 \times g_T(\mu = m_b) \, .
\eea
Furthermore, we included the small electroweak corrections discussed in Ref.~\cite{Gonzalez-Alonso:2017iyc} which can induce mixing between the scalar and tensor contributions. 

\begin{figure}[t!]
\centering
\includegraphics[width=0.7\linewidth]{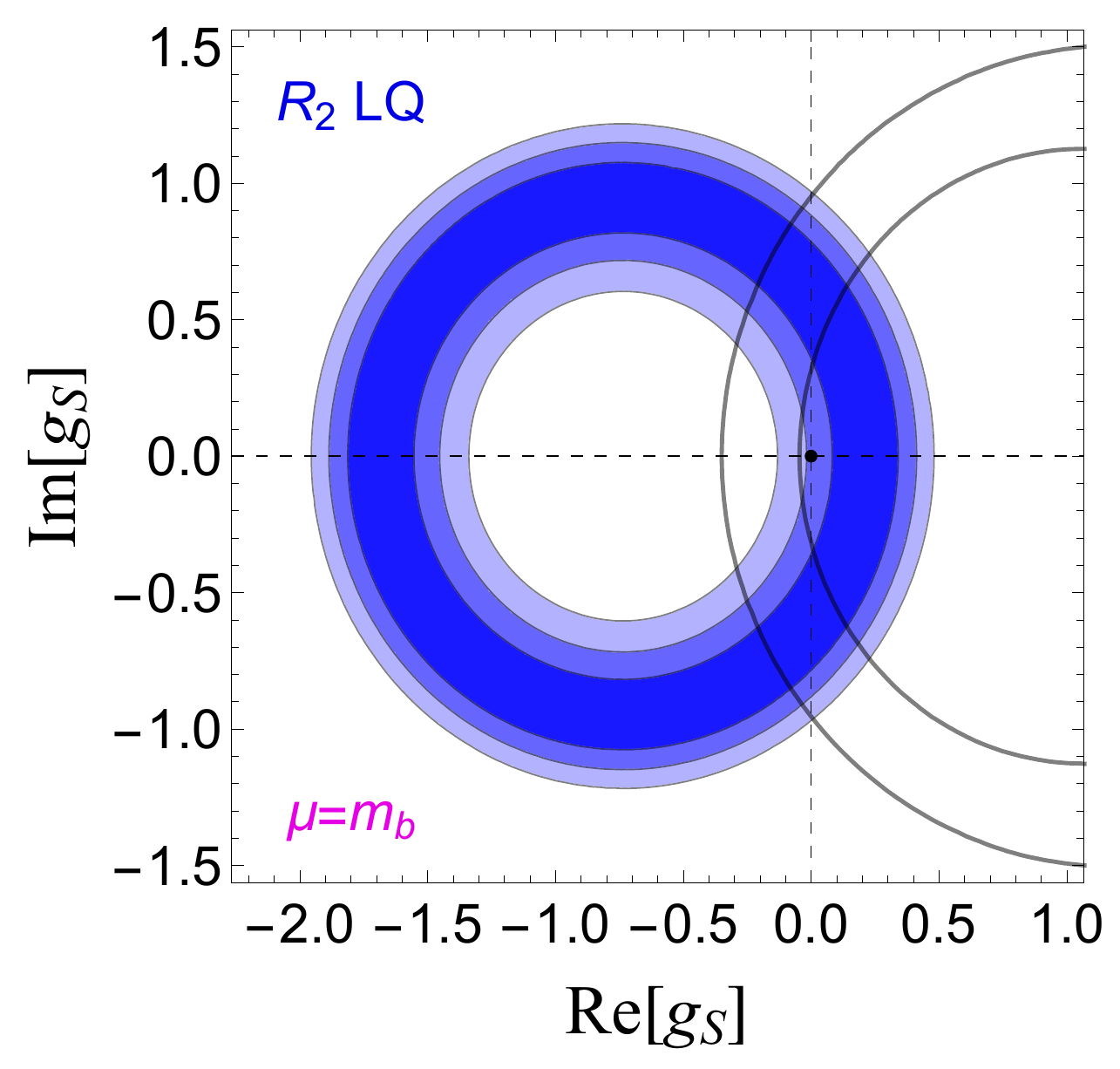}
\caption{\small \sl Regions of allowed values for $g_S(\mu=m_b)$ on the complex plane, compatible with experimentally measured $R_D$ to $1-$, $2-$ and $3\sigma$ (from darker to lighter blue). We assumed $g_S(\mu=m_\Delta)=4 \times g_T(\mu=m_\Delta)$, as predicted by the $R_2$ LQ scenario, and considered the QCD running from $\mu\approx1~\mathrm{TeV}$ to $m_b$, cf.~Eq.~\eqref{eq:running}. We also show in black the region compatible to $2\sigma$ accuracy with $R_{D^\ast}^{\mathrm{exp}}$, which indicates that the effective couplings must be complex in this scenario. See discussion in the text.}
\label{fig:RD-fit}
\end{figure}

 We focus on $B\to D$ semileptonic decays for which the hadronic form factors have been computed by means of numerical simulations of QCD on the lattice in Refs.~\cite{Na:2015kha,Lattice:2015rga}.~\footnote{For the reader's convenience we provide in Appendix~\ref{app:flavor}  the expression for the $B\to D \ell \bar{\nu}$ decay rate.} To fill the discrepancy between $R_D^{\rm SM}$ and $R_D^{\rm exp}$ one then allows for $g_S\neq 0$. In this way we find the region of allowed values of $\mathrm{Re}[g_S(m_b)]$ and $\mathrm{Im}[g_S(m_b)]$ which is depicted by the blue regions in Fig.~\ref{fig:RD-fit} to $1$, $2$ and $3\sigma$ accuracy. 
We decided not to include $R_{D^\ast}$ in the same fit because the lattice QCD determination of the form factors at non-zero recoil has not been made, in addition to the fact that the relevant pseudoscalar form factor has never been computed on the lattice. To verify that our solution to $R_D$ also gives an improvement of $R_{D^\ast}$, despite the uncertainties mentioned above which need to be clarified by lattice simulations, we show in the same plot the $2\sigma$ region consistent with $R_{D^\ast}^{\mathrm{exp}}$. The latter region was obtained by using the $B\to D^\ast \ell \nu_\ell$ form factors extracted from experimental results~\cite{Amhis:2016xyh}, combined with the ratio $A_0(q^2)/A_1(q^2)$ and $T_{1-3}(q^2)/A_{1}(q^2)$ computed in Ref.~\cite{Bernlochner:2017jka}. By using this approach, we conclude that, in the $R_2$-model, only the complex $g_S$ solutions to $R_D$ can provide a reasonable agreement with $R_{D^{(\ast)}}^{\rm exp}$, a fact that agrees with findings of Ref.~\cite{Sakaki:2013bfa}. Therefore, we will focus our analysis on the solutions to $R_D$ with complex Wilson coefficients
\begin{equation}
|g_{S}(\mu=m_b)| \in (0.56,0.74)\,,
\end{equation}
obtained to $1\sigma$ accuracy. This relation implies that the magnitude of LQ couplings should be such that
\begin{equation}
\dfrac{|g_L^{c\tau}| |g_R^{b\tau}|}{m_\Delta^2} \in (0.80,1.32) \times (1~\mathrm{TeV})^{-2}\,,
\end{equation}
\noindent where we used $|V_{cb}|=0.0417(20)$~\cite{Bona:2006ah}. 
We see that the couplings needed to explain $R_D$ for $m_\Delta\lesssim 1~\mathrm{TeV}$ should not be too large, mostly due to the CKM enhancement in Eq.~\eqref{eq:gsgt}.

\

In the following we show that the explanation of $R_{D^{(\ast)}}^{\mathrm{exp}}>R_{D^{(\ast)}}^{\mathrm{SM}}$ described above is consistent with other limits from flavor physics and the direct searches for pair-produced LQs at the LHC. To that purpose, we will assume that the coupling $g_R^{b\tau}$ is purely imaginary.

\subsection{General Constraints}
\label{sec:cons}

The choice of $g_L^{c\tau}\neq 0$ and $g_R^{b\tau}\neq 0$ in
Eq.~\eqref{eq:YC} allows us to avoid many flavor physics constraints,
making this explanation of $R_{D^{(\ast)}}^{\mathrm{exp}}>R_{D^{(\ast)}}^{\mathrm{SM}}$
particularly simple. Processes mediated by the flavor-changing neutral $\Delta
F=1$ and $\Delta F=2$ currents are not altered in this
scenario. Only the charged-current transitions $b\to
c\tau\bar{\nu}$ are modified, as desired.  The most significant constraints to this
scenario actually come from the $Z$ boson decay widths.

\subsubsection{Constraints from $Z$ decays}
\begin{figure}[t!]
\centering
\includegraphics[width=0.25\linewidth]{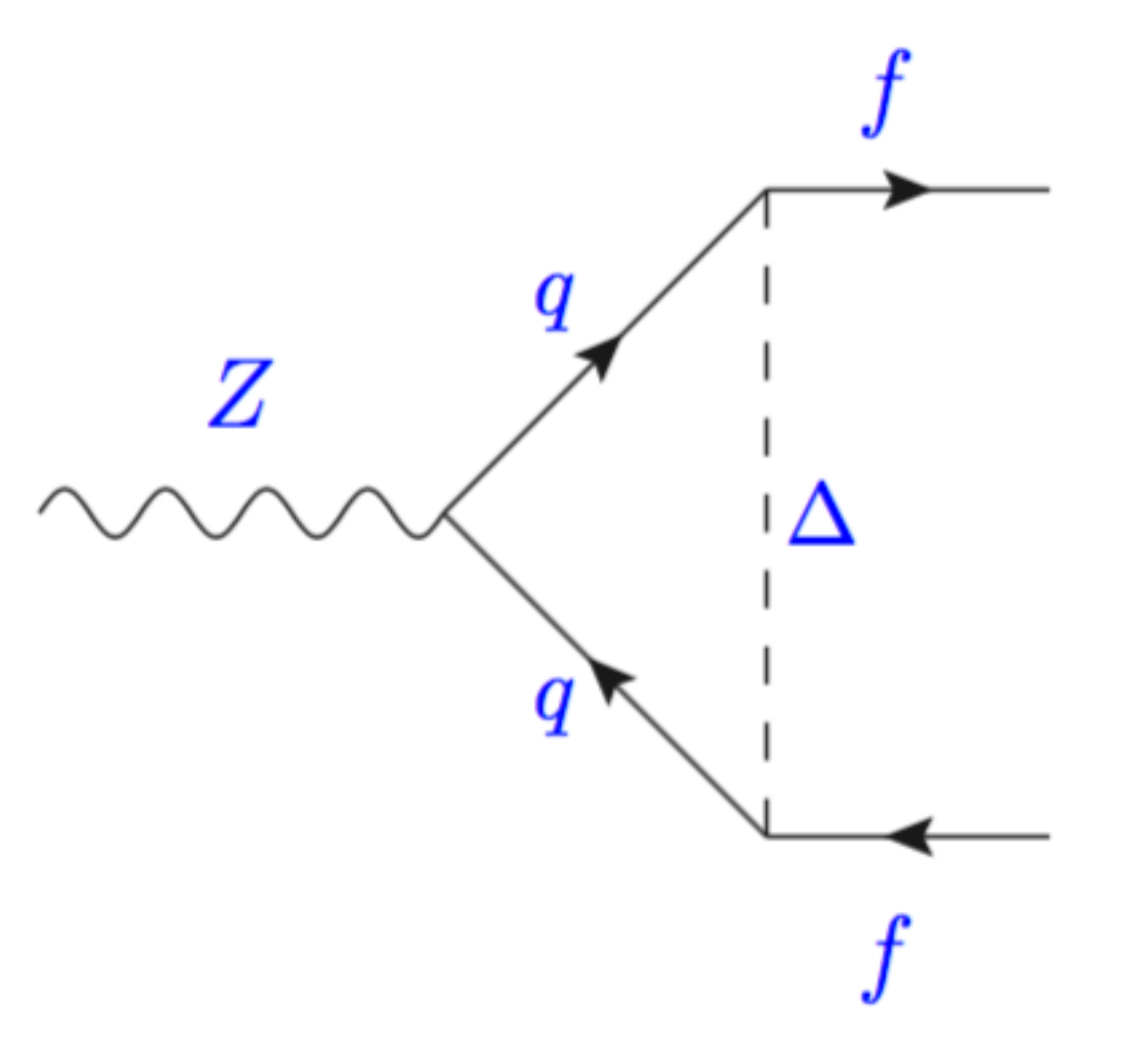}~\includegraphics[width=0.25\linewidth]{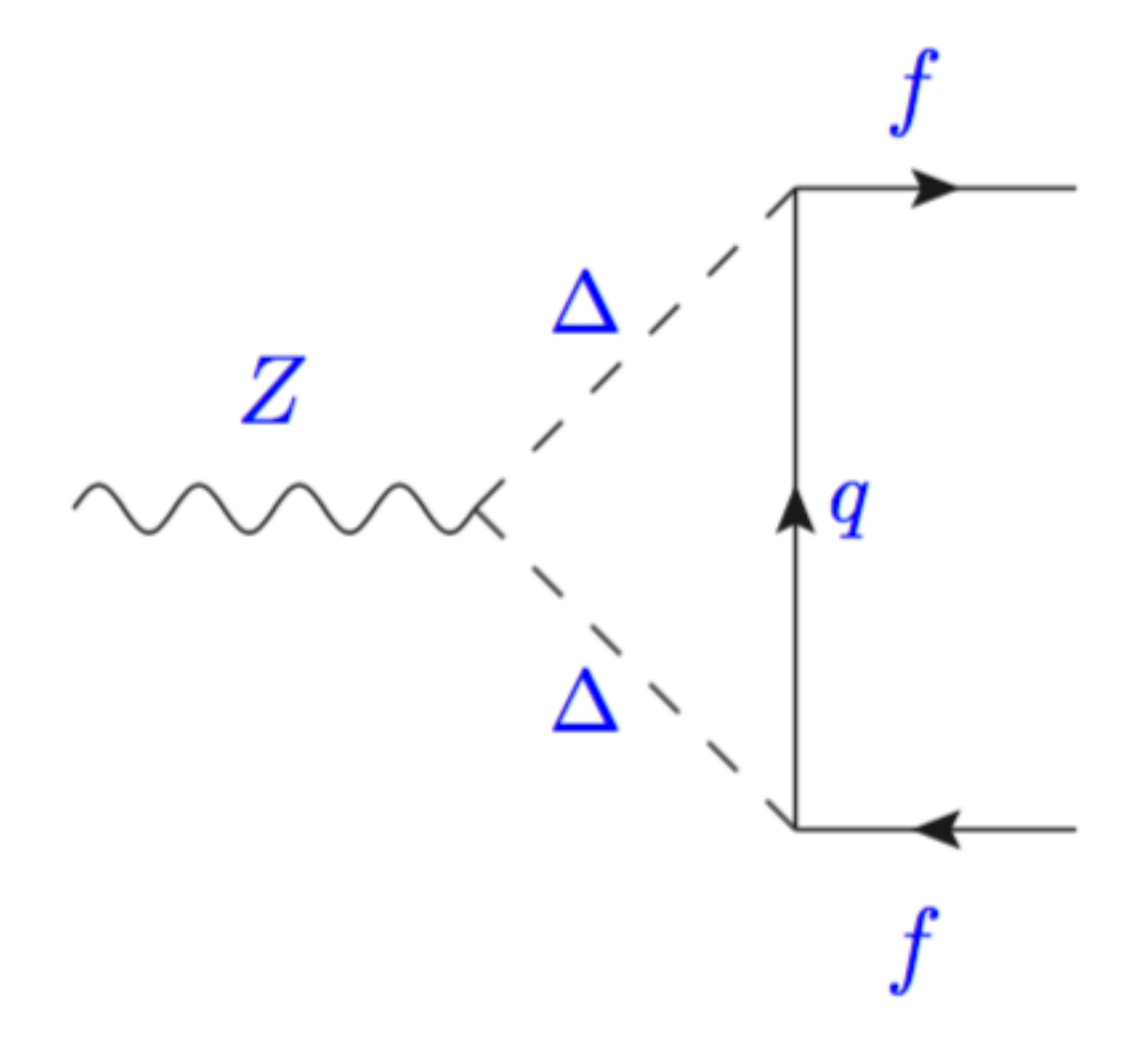}~\includegraphics[width=0.25\linewidth]{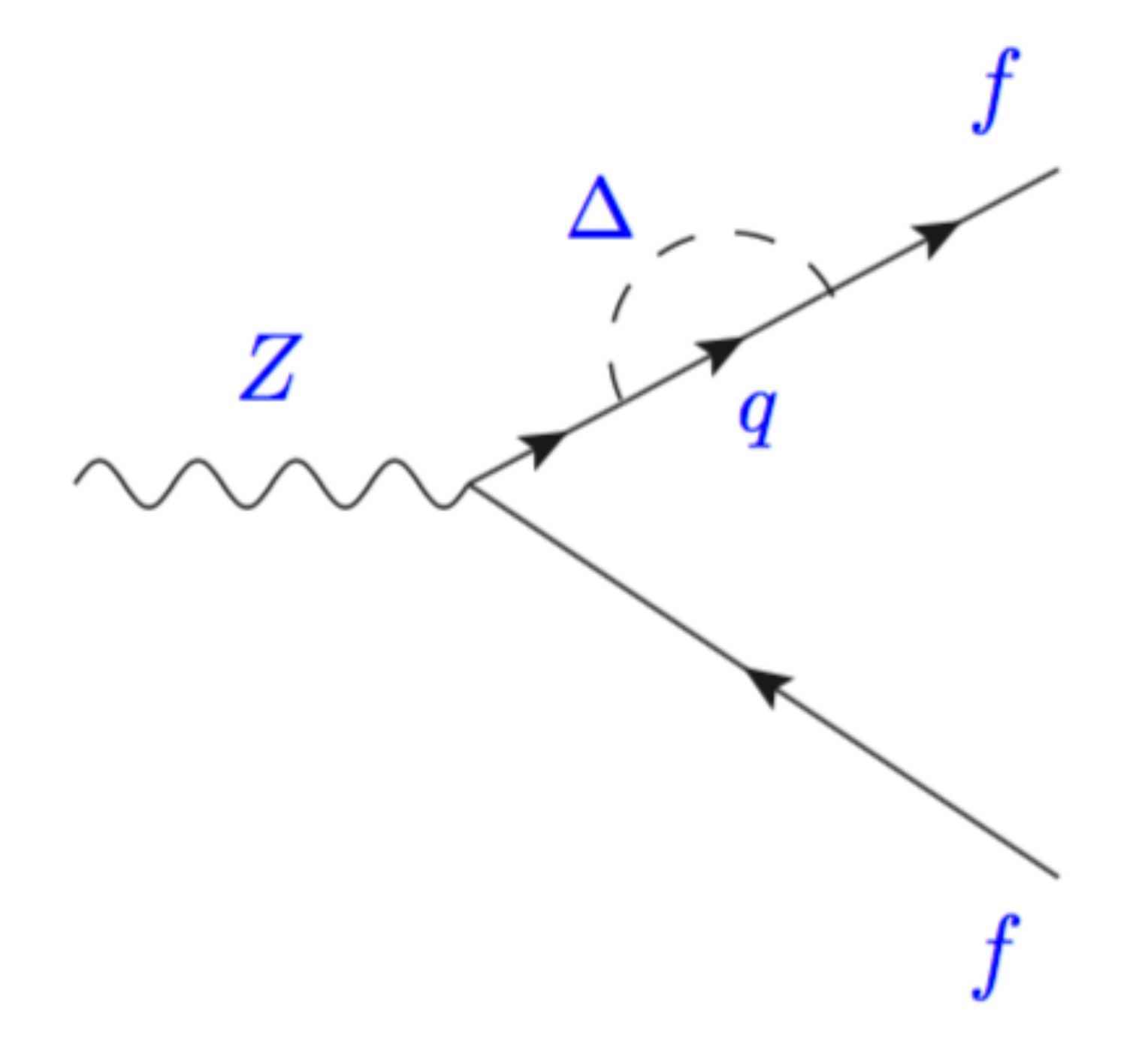}~\includegraphics[width=0.25\linewidth]{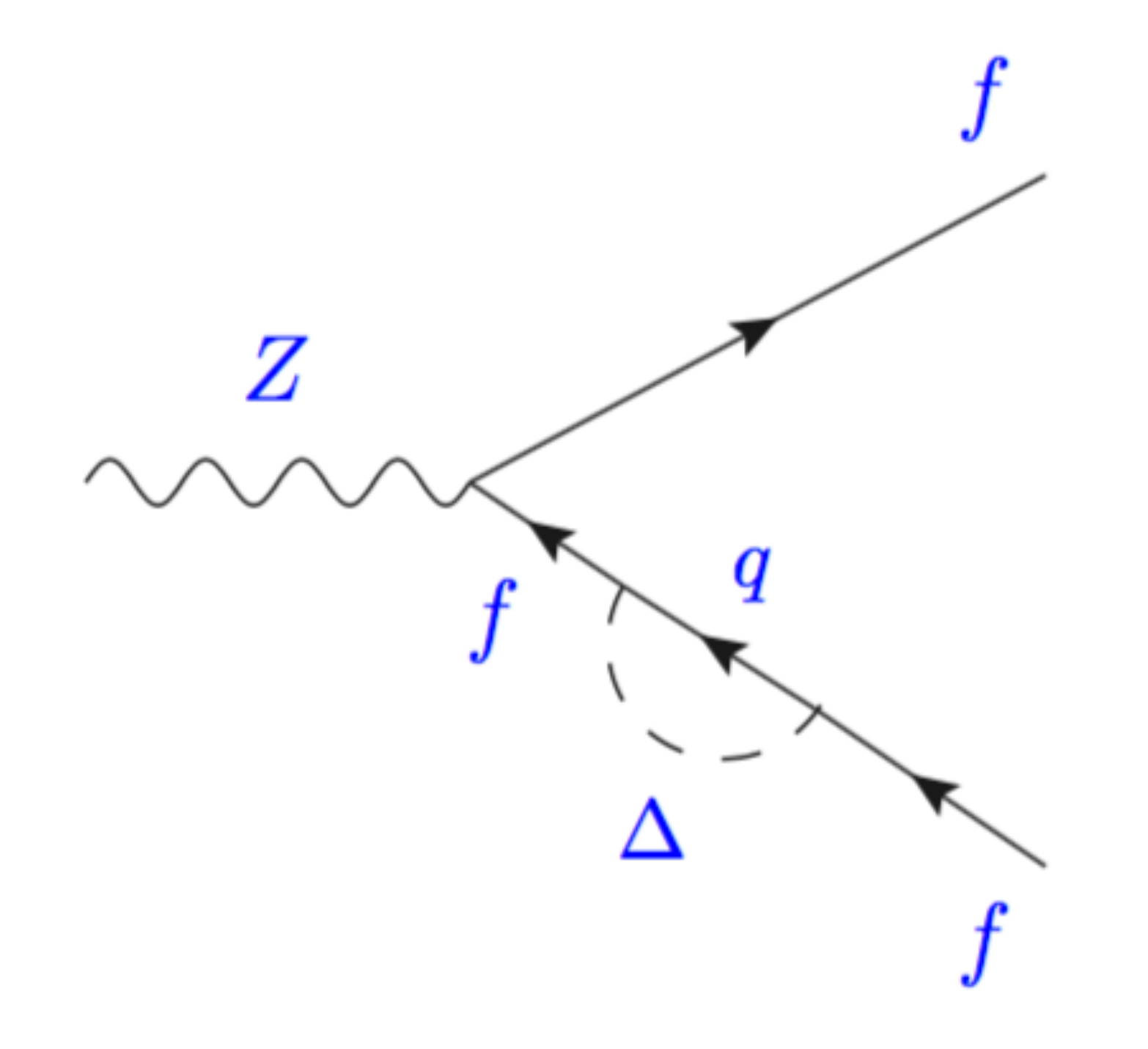}
\caption{\small \sl Contributions to the $Z\to f \bar f$ decay amplitude generated in our $R_2$ model, where the LQ $\Delta$ is either $\Delta^{(2/3)}$ or $\Delta^{(5/3)}$, $q\in\{ u,c,b,t\}$ 
depending on the LQ state running in the loop, while $f$ is either $\ell$ or $\nu$. Similar diagrams can be drawn for $Z\to b\bar{b}$.}
\label{fig:Zdecays}
\end{figure}

Our LQ model induces modifications to $Z$-boson decays
through loop contributions, as depicted in Fig.~\ref{fig:Zdecays}. Those effects have to be consistent with the LEP measurements of $\mathcal{B}(Z\to \ell^+\ell^-)$~\cite{Patrignani:2016xqp}. The Lagrangian describing the $Zf\bar f$ coupling can be written as
\bea {\cal L}_{Z f\bar{f}} = \frac{g}{\cos \theta_W} Z^\mu \bar{f}
\gamma_\mu \left ( g_L^f \,P_L + g_R^f \,P_R \right )f \, ,
\label{eq:Zff}
\eea
where $g$ is the $SU(2)_L$ gauge coupling, and
\begin{align}
\begin{split}
 g_L^f &= g_L^{f,\,\mathrm{SM}} + \delta g_L^f\,,\\
 g_R^f &= g_R^{f,\,\mathrm{SM}} + \delta g_R^f\,,
\end{split}
\end{align}
where $g_{L}^{f,\,\mathrm{SM}} = T_3-Q \sin^2 \theta_W$ and  $g_{R}^{f,\,\mathrm{SM}} = -Q \sin^2 \theta_W$  
are the SM contributions, while 
$\delta g_{L,R}^f$ parametrize LQ loop contributions. The corresponding branching ratio is then given by
\begin{align}
\label{eq:ZBR}
\begin{split}
{\cal B}(Z \to f \bar f) = \frac{m_Z^3}{6 \pi v^2 \Gamma_Z}
\, \beta_f &\,\left[ \left[\big{|}g_L^{f}\big{|}^2+\big{|}g_R^f\big{|}^2\right]\left(1-\dfrac{m_f^2}{m_Z^2}\right)+ \dfrac{6 m_f^2}{m_Z^2}\mathrm{Re}\left(g_L^f \, g_R^{f\ast}\right)\, \right],
\end{split}
\end{align}
where $m_f$ is the fermion mass and $\beta_f=\sqrt{1-4 m_f^2/m_Z^2}$. 

The LQ couplings lead to up- and down-type quark loop contributions to $\delta g_{L(R)}^{\tau}$, which are given by
\begin{align}
\label{eq:Zcoupl-gRtau}
\begin{split}
\delta g_R^\tau &= N_c \,\vert g_R^{b\tau}\vert^2 \Bigg{\lbrace}\vert V_{tb}\vert^2 \,\frac{x_t}{32 \pi^2} 
\left( 1+ \log x_t \right)\\
&+\dfrac{x_z}{144 \pi^2}\Bigg{[}-\left((\sin^2\theta_W-\dfrac{3}{2}\right)\left(\log x_z + i \pi\right)+\left(-\dfrac{1}{4}+\dfrac{2}{3}\sin^2\theta_W\right)\Bigg{]}\\
&+(1- \vert V_{tb}\vert^2)\,\dfrac{x_z}{72 \pi^2}  \Bigg{[}\left(\sin^2\theta_W - \frac{3}{4}\right)\left(\log x_z+ i \pi +\dfrac{1}{12}\right)+\dfrac{3}{16}\Bigg{]}\Bigg{\rbrace}\,,
\end{split}
\end{align}
and
\begin{align}
\label{eq:Zcoupl-gLtau}
\begin{split}
\delta  g_L^\tau &= N_c\,\vert g_L^{c\tau}\vert^2\dfrac{x_z}{72\pi^2}  \Bigg{[}\sin^2\theta_W \left(\log x_z+i\pi+\dfrac{1}{12}\right)-\dfrac{1}{8}\Bigg{]},
\end{split}
\end{align}
where $x_t =m^2_t/m^2_\Delta$ and $x_z =m^2_Z/m^2_\Delta$, and $N_c=3$
is the number of colors.~\footnote{One can easily identify the top and
  bottom quark contributions in the first two lines of
  Eq.~\eqref{eq:Zcoupl-gRtau}, while the ones coming from the up and charm quarks   have been recast in the same equation by using the unitarity of
  $V_{\mathrm{CKM}}$, i.e.~$|V_{ub} \, g_{R}^{b\tau}|^2+|V_{cb} \,
  g_{R}^{b\tau}|^2=|g_R^{b\tau}|^2\left(1-|V_{tb}|^2\right)$.} The
dominant term in $\delta g_R^\tau$ comes from the top quark since it
enhances the loop function. The effective coupling $\delta g_L^\tau$
does not exhibit a similar mass enhancement, and it is for this reason
less relevant for phenomenology. Similarly, the LQ interaction with
the charm quarks also induces an effective coupling to $\tau$
neutrinos, which reads

\begin{equation}
\delta g_L^{\nu_\tau} = N_c\,\vert g_L^{c\tau}\vert^2 \dfrac{x_z}{72 \pi^2}\Bigg{[}\sin^2\theta_W \left(\log x_z + i\pi -\dfrac{1}{6}\right)+\dfrac{1}{8}\Bigg{]}\,. 
\end{equation}

\noindent The effective couplings given above are constrained by the LEP measurements of the $Z$ decay widths and other electroweak observables~\cite{Patrignani:2016xqp}. The measurements of the effective $Z$ couplings give~\cite{ALEPH:2005ab},
\begin{equation}
\dfrac{g_V^\tau}{g_V^e} = 0.959(29)\,,\qquad\quad\qquad \dfrac{g_A^\tau}{g_A^e} = 1.0019(15)\,,
\end{equation}

\noindent where $g_{V(A)}^f = g_L^f \pm g_R^f$. These results translate into useful bounds on the LQ contributions to $\delta g_{L,R}^\tau$. 
Furthermore, the LEP bound on the effective number of neutrinos~\cite{ALEPH:2005ab}
\begin{equation}
N_\nu = 2.9840(82)\,,
\end{equation}

\noindent  will constrain the LQ couplings via the expression,

\begin{equation}
N_\nu = 2 + \left\vert 1+\dfrac{\delta g_L^{\nu_\tau}}{g_L^{\nu_\tau,\,\mathrm{SM}}}\right\vert^2\,.
\end{equation}

\noindent As we shall see below, the coupling $g_R^{b\tau}$ is tightly limited by $\mathcal{B}(Z\to \tau\tau)$ due to the enhancement of the loop function by the top quark mass. On the other hand, the constraints on $g_L^{c\tau}$ derived from $N_\nu$ would be relevant only for very light LQ states, already excluded by the constraints stemming from $g_{V,A}^\tau$. Finally, we have also checked that the bound on the LQ couplings arising from 
$\mathcal{B}(Z\to b\bar{b})$ is not significant with current experimental precision.

\subsubsection{Limits from Direct Detection}
Direct searches at the LHC for pair-produced LQs provide useful limits on the couplings  $g_L^{c\tau}$ and $g_R^{b\tau}$ as a function of the LQ mass $m_\Delta$. For the flavor ansatz given in Eq.~\eqref{eq:YC} the allowed LQ decay modes are
\begin{align}
\begin{split}
\Delta^{(2/3)} &\to c\nu,~b\tau\,,\\
\Delta^{(5/3)} &\to u\tau,~c\tau,~t\tau\,.
\end{split}
\end{align}
The CMS collaboration recently improved bounds of LQs decaying to $b\tau$, obtaining a lower bound on $m_\Delta$ of $900$ GeV, with the assumption ${\cal B}(\Delta^{(2/3)} \to b \tau)=1$~\cite{Khachatryan:2016jqo,CMS:2016hsa}. Similarly, CMS excludes LQs decaying into $t\tau$ with masses $m_\Delta \gtrsim 850~\mathrm{GeV}$ if $\mathcal{B}(\Delta^{(5/3)}\to t\tau)=1$~\cite{Khachatryan:2015bsa,Sirunyan:2018nkj}. Furthermore, a search for pair-produced LQs decaying into a light quark ($q=u,d,s,c$) and a neutrino has been reported by CMS, allowing us to set the limit $m_\Delta\gtrsim 980$~GeV if  $\mathcal{B}(\Delta^{(2/3)}\to c\nu)=1$~\cite{CMS:2018bhq}.

\begin{figure}[htp!]
\centering
\includegraphics[width=0.5\linewidth]{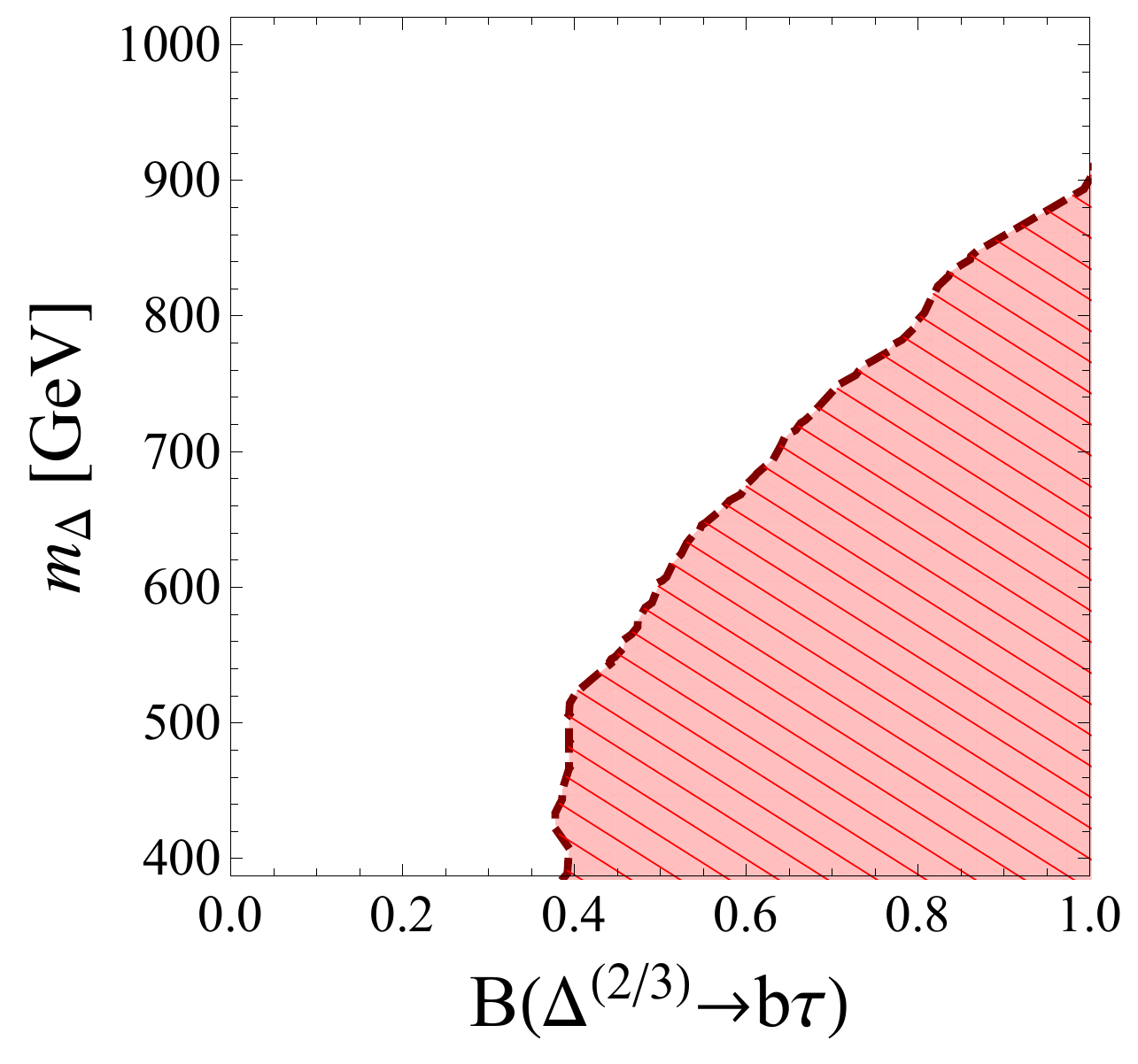}~\includegraphics[width=0.5\linewidth]{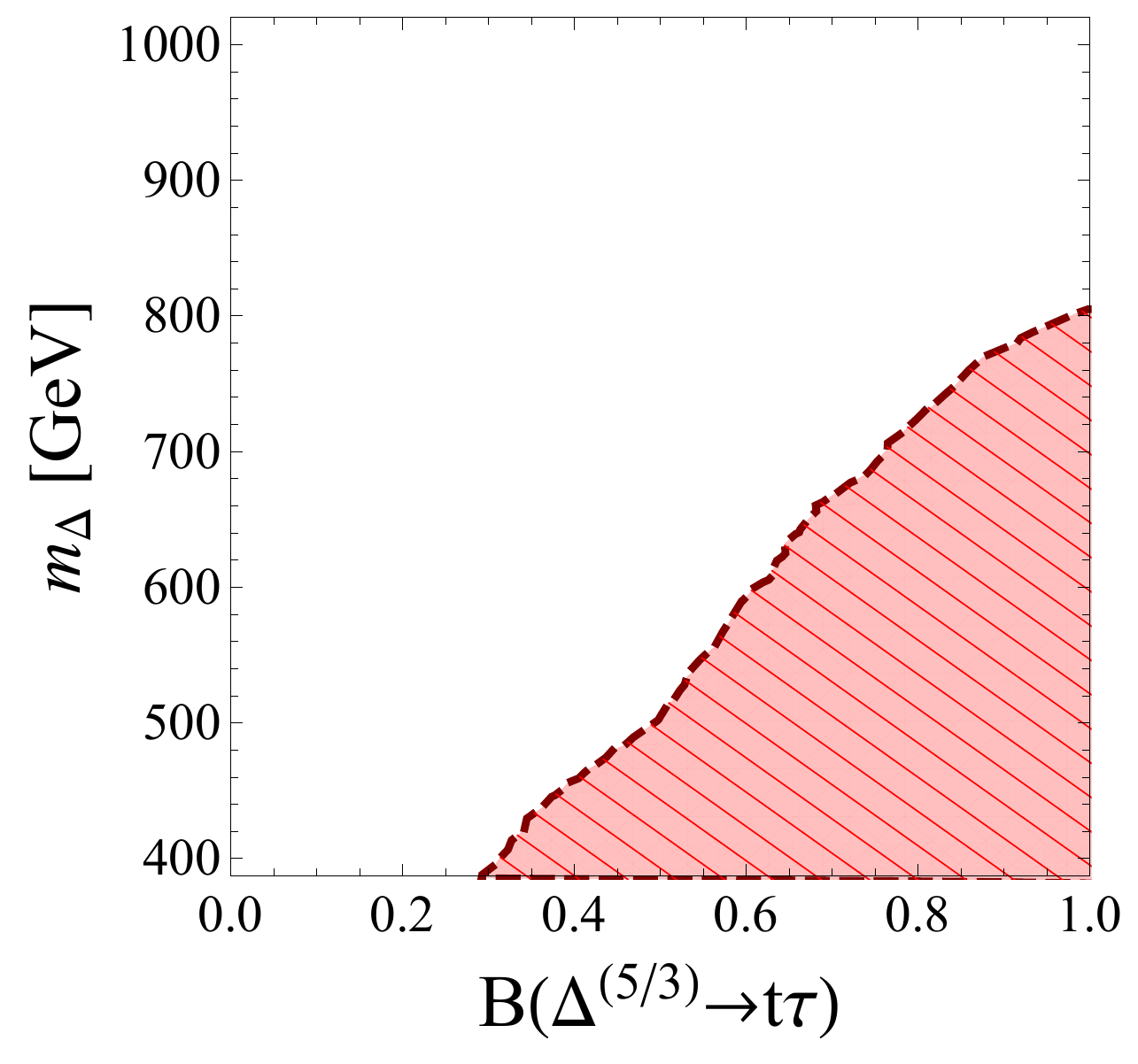}\\[0.4em]
\includegraphics[width=0.5\linewidth]{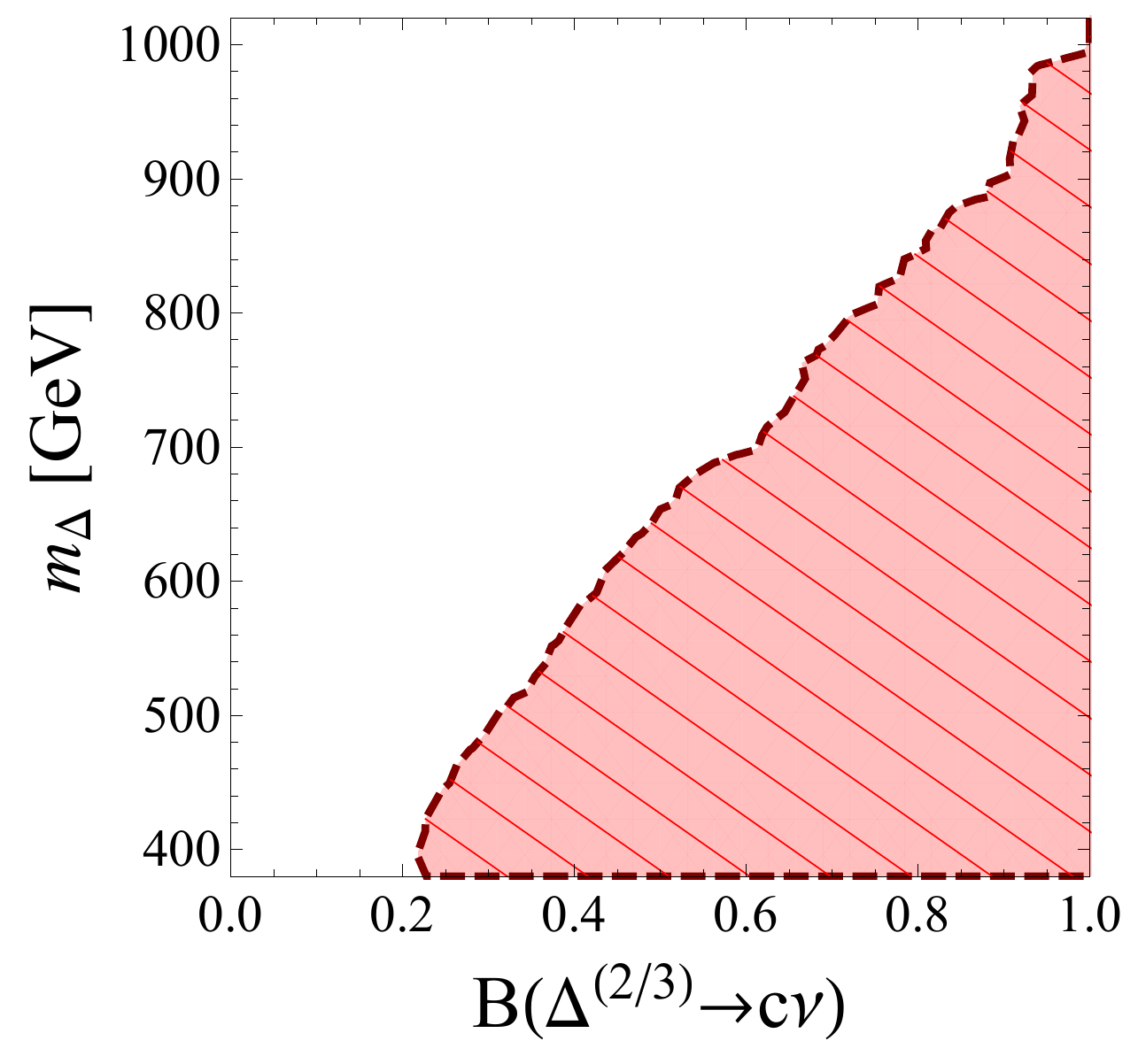}
\caption{\small \sl Exclusion regions of $m_\Delta$ as a function of $\mathcal{B}(\Delta^{(2/3)}\to b\tau)$~\cite{Khachatryan:2016jqo,CMS:2016hsa}, $\mathcal{B}(\Delta^{(5/3)}\to t\tau)$~\cite{Khachatryan:2015bsa,Sirunyan:2018nkj} and  $\mathcal{B}(\Delta^{(2/3)}\to c\nu)$~\cite{CMS:2018bhq} at $95\%$~CL. See text for details.}
\label{fig:direct-searches}
\end{figure}

In any realistic scenario, the limits mentioned above have to be reinterpreted to account for values of the branching ratios smaller than one. The reinterpretation of these exclusion limits on $m_\Delta$ are shown in Fig.~\ref{fig:direct-searches} as a function of ${\cal B}(\Delta^{(2/3)} \to b \tau)$ (upper left panel), ${\cal B}(\Delta^{(5/3)} \to t \tau)$ (upper right panel) and ${\cal B}(\Delta^{(2/3)} \to c \nu)$ (lower panel). In the scenario we are considering, cf. Eq.~\eqref{eq:YC}, the branching ratios read
\begin{align}
\mathcal{B}(\Delta^{(2/3)}\to c\nu)  &= \dfrac{|g_L^{c\tau}|^2}{|g_R^{b\tau}|^2+|g_L^{c\tau}|^2}\,,\qquad\qquad
\mathcal{B}(\Delta^{(2/3)}\to b\tau) = \dfrac{|g_R^{b\tau}|^2}{|g_R^{b\tau}|^2+|g_L^{c\tau}|^2}\,,
\end{align}
and
\begin{align}
\mathcal{B}(\Delta^{(5/3)}\to t\tau) &= \dfrac{|g_R^{b\tau}|^2\left(1-\dfrac{m_t^2}{m_\Delta^2}\right)^2}{|g_R^{b\tau}|^2\left(1-\dfrac{m_t^2}{m_\Delta^2}\right)^2+|g_L^{c\tau}|^2}\,,
\end{align}
where, for simplicity, we assumed $|V_{tb}|\approx 1$ and neglected the light fermion masses ($m_{c}$, $m_b$ and $m_\tau$). From the above
formulas we conclude that scenarios with $m_\Delta \lesssim 1$~TeV can comply with the
limits of Fig.~\ref{fig:direct-searches} if the branching ratios are
diluted by a large coupling to the charm quark. In other words, the
ratio $|g_R^{b\tau}|/|g_L^{c\tau}|$ should be of 
  $\mathcal{O}(1)$, which also agrees with the
constraints on $g_R^{b\tau}$ derived from $Z\to\tau^+\tau^-$,
cf.~Eq.~\eqref{eq:Zcoupl-gRtau}. We checked that $m_\Delta$
can be as low as $\approx 650$~GeV for some combinations of Yukawa couplings.

\begin{figure}[htp!]
\begin{center}
\includegraphics[width=0.5\linewidth]{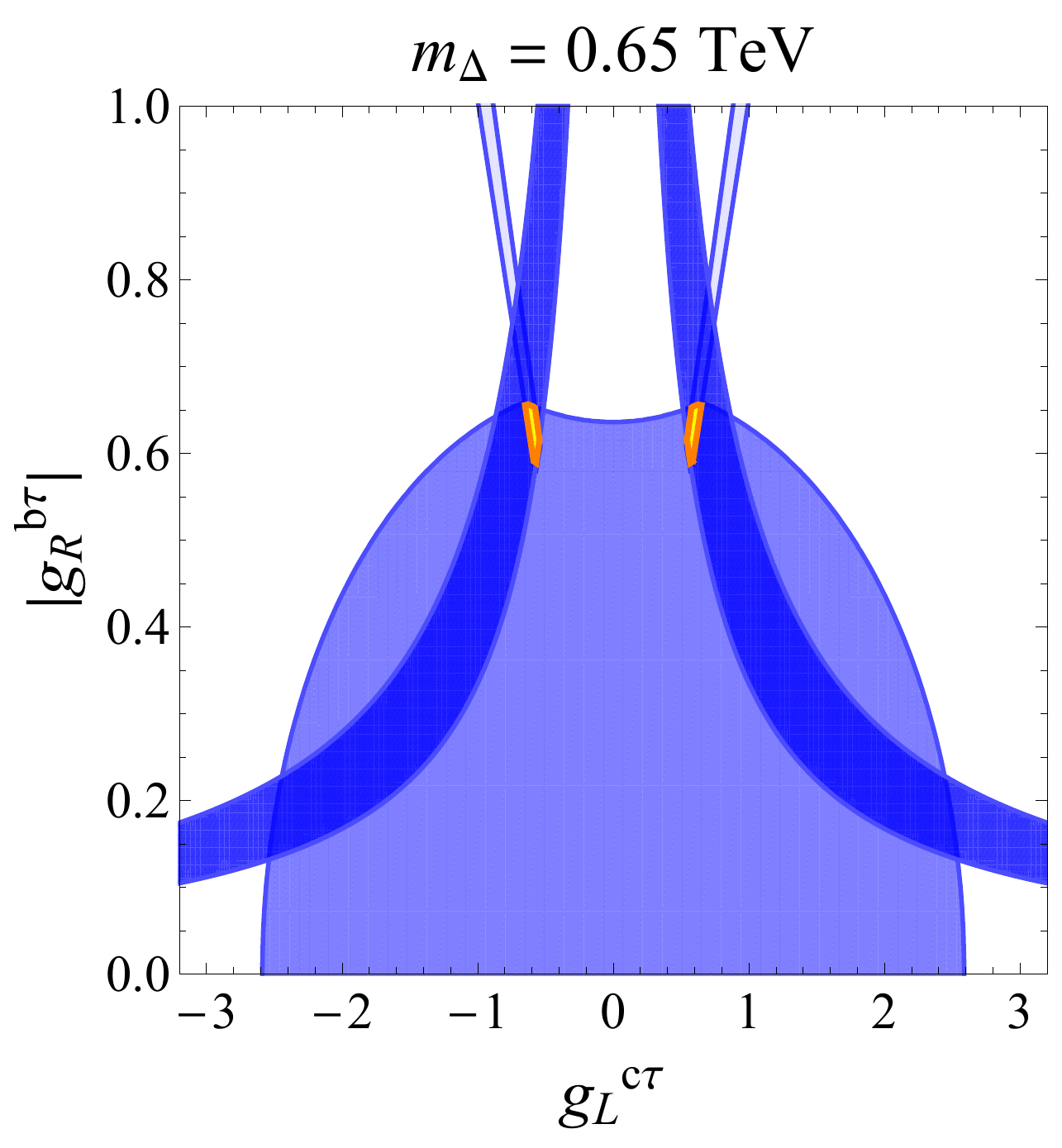}~\includegraphics[width=0.5\linewidth]{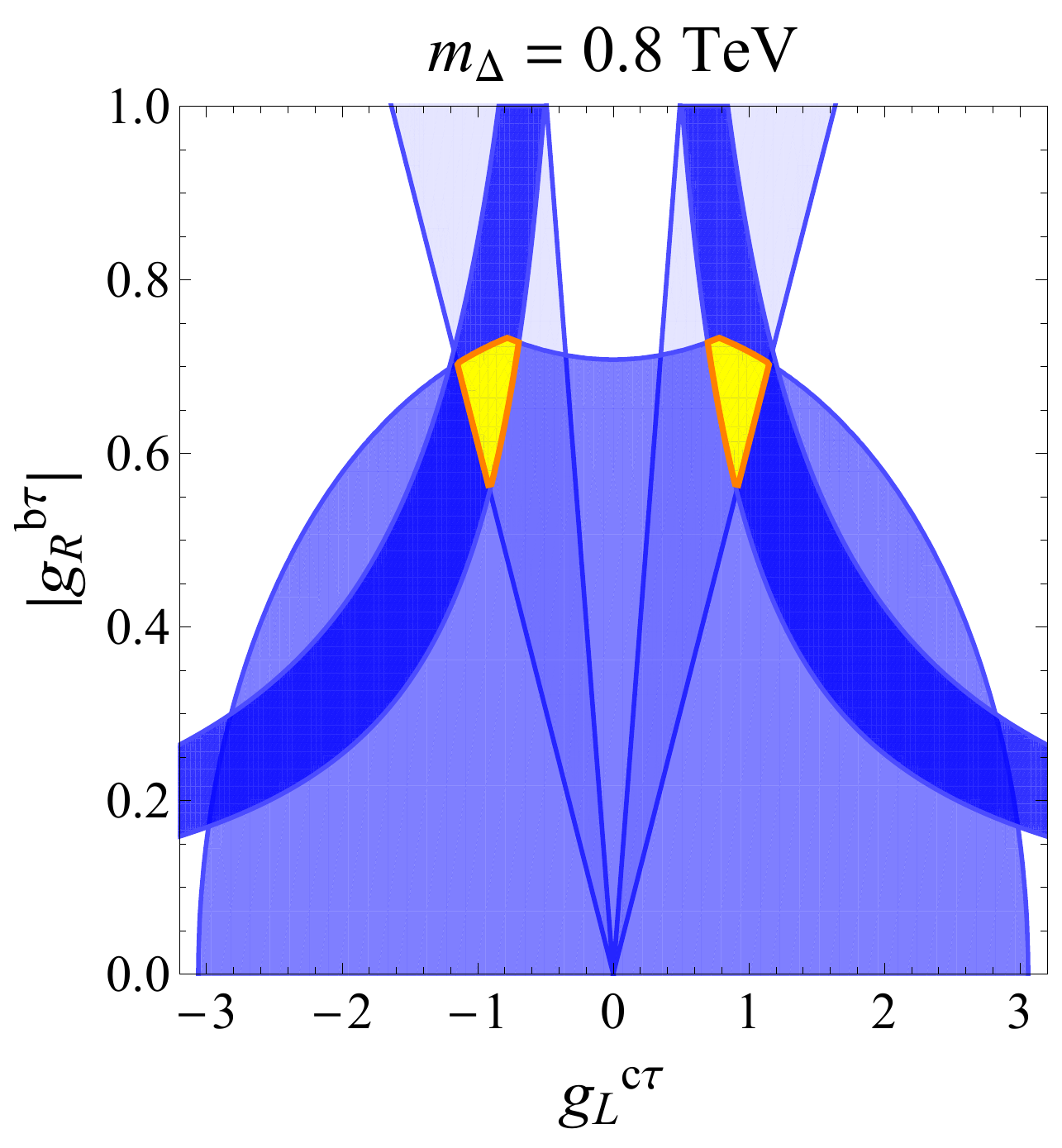}\\\includegraphics[width=0.5\linewidth]{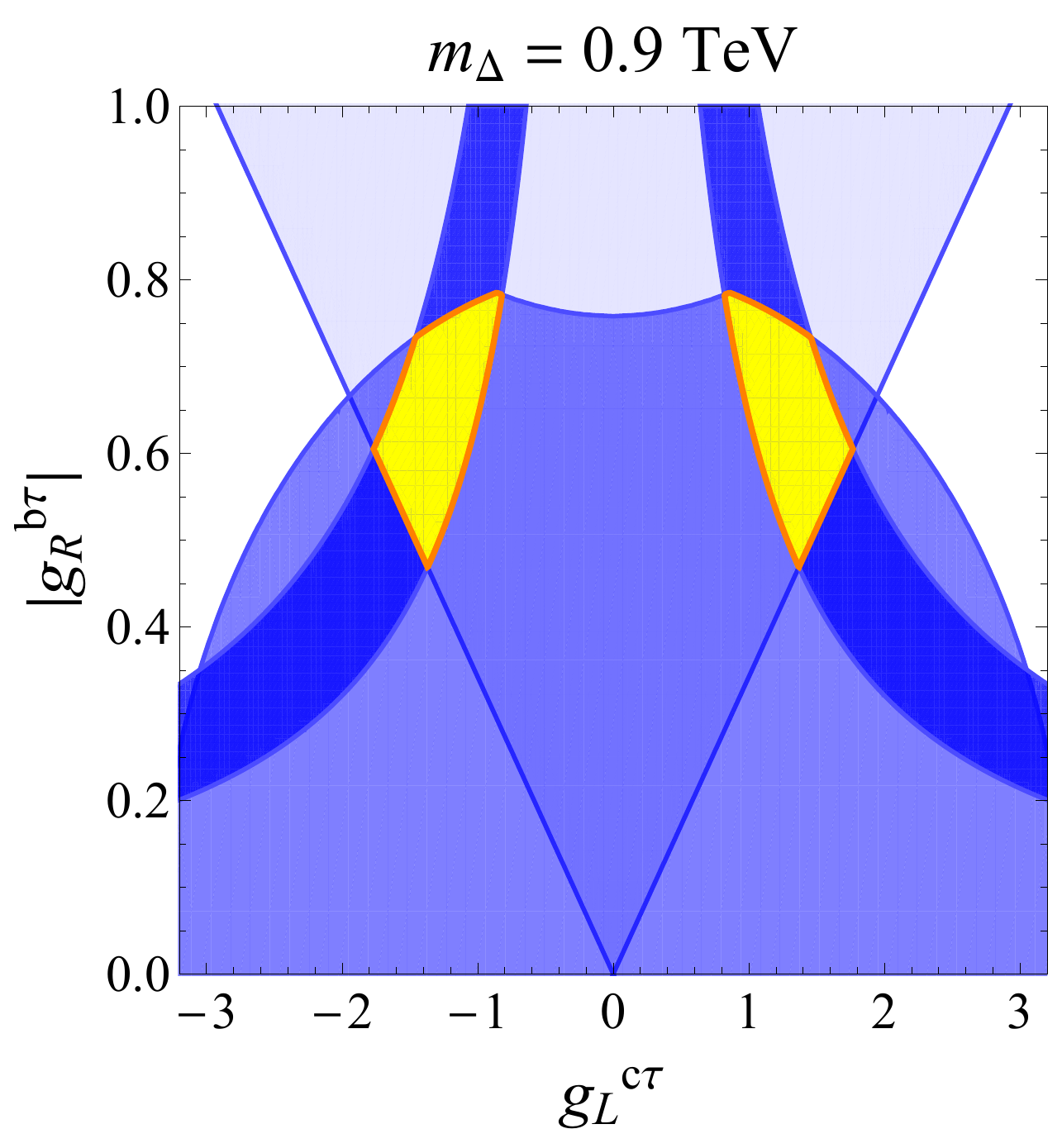}
\end{center}
\caption{The allowed region on the plane $g_L^{c\tau} \times
  |g_R^{b\tau}|$ by current experimental constraints is shown in
  yellow for $m_\Delta\in \lbrace 0.65,~0.8,~0.9\rbrace$~TeV. We
  assume $g_R^{b\tau}$ to be a purely imaginary coupling, as described
  in the text. The separate allowed regions both from the direct
  searches at the LHC, the $Z$-pole observables and $R_D$ are shown
  from lighter to darker blue, respectively. Thus, the yellow region
  is given by their intersection. See text for details. }
\label{fig:flavorlimits2}
\end{figure}

In Fig.~\ref{fig:flavorlimits2}, we confront the allowed region by
$R_D$ with the indirect and direct constraints discussed above by
assuming $g_R^{b\tau}$ to be purely imaginary, as discussed in Sec.~\ref{sec:RD}. For instance, we see in
these plots that both $|g_R^{b\tau}|$ and $g_L^{c\tau}$ are bounded
from above by the $Z$-pole observables. Furthermore, the couplings needed to explain $R_D$ for
$m_\Delta\lesssim 1~\mathrm{TeV}$ are not very large, mostly due to
the CKM enhancement in Eq.~\eqref{eq:gsgt}, and they can therefore be
perfectly consistent with phenomenological constraints discussed in
this Section.

\

In what follows, we will discuss the IceCube phenomenology of the
LQ scenario considered in this paper. In practice, we will
focus on the analysis of the parameter space defined by $m_\Delta$ and
$g_L^{c\tau}$ since they are the relevant parameters to determine
IceCube signals. In this plane the region allowed by the $Z$-pole
observables is approximately limited by a straight line, such that the
allowed values are $g_L^{c\tau} \lesssim 1.6
+1.3(m_\Delta/300\,\text{GeV} - 1)$, which is also illustrated in this
paper in Figs.~\ref{fig:StatisticalTestCurrentData} and
\ref{fig:Projections}. 


\section{\label{sec:3} High Energy Neutrino Events in IceCube}

The cubic kilometer IceCube Neutrino Observatory in the South Pole
observed between 2010 and 2012 the first evidence of a diffuse flux of
astrophysical neutrinos above 100 TeV~\cite{Aartsen:2013jdh}, which
was further confirmed in the updated analyses~\cite{Aartsen:2014gkd,Aartsen:2015zva,Aartsen:2017mau}. In this paper
we consider the 6-year sample~\cite{Aartsen:2017mau} which contains 80 High Energy Starting
Events (HESE). In particular, we focus on the 28 events with deposited energies above 100 TeV, which are
divided in 23 showers and 5 tracks. This flux, of yet unknown astrophysical
origin, seems to be isotropic and consistent with a single power law
behavior followed by neutrinos and antineutrinos of all
flavors~\cite{Aartsen:2015ivb}. In other words, the neutrino energy spectrum can be
parameterized by a spectral index, $\gamma$, and a normalization
constant $\rm C_0$ as 
\bea 
\frac{d\Phi_\nu}{dE_\nu} = \frac{\rm C_0}{10^{8}}
\left( \frac{E_\nu}{100 \; \rm TeV}\right)^{2-\gamma} \frac{1}{E_\nu^2}\quad {\rm
  [GeV^{-1} \; cm^{-2} \; str^{-1} \; s^{-1}]}\,, 
\label{inspectrum}
\eea 
so that at the Earth the flavor fluxes are given by
\begin{equation}
\frac{d\Phi_{\alpha}}{dE_\nu} =
  f^\oplus_{\alpha} \; \frac{d\Phi_\nu}{dE_\nu} \,,
\end{equation}
where $f^\oplus_{\alpha}$,  $\alpha=\{\nu_e,\nu_\mu,\nu_\tau,\bar \nu_e,\bar \nu_\mu,\bar \nu_\tau  \}$, are the normalized neutrino and antineutrino flavor fractions at the Earth. 

\begin{figure}[htb]
\begin{center}
\begin{minipage}[b]{0.48\textwidth}
\includegraphics[width=\textwidth]{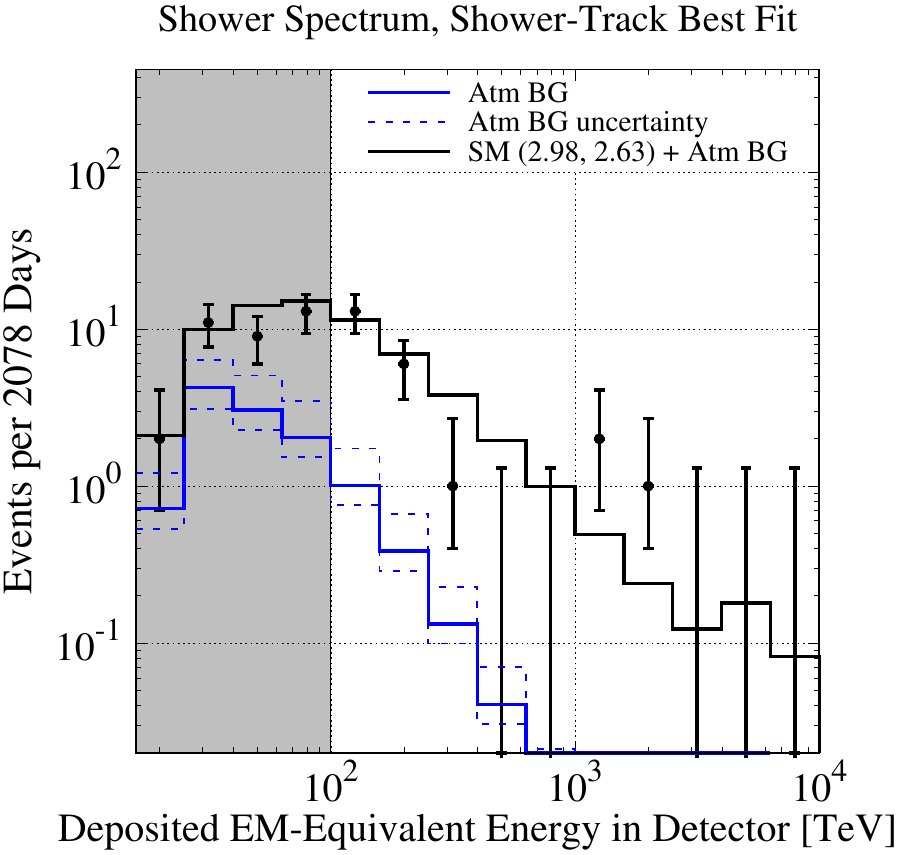}
\end{minipage}
\hfill
\begin{minipage}[b]{0.48\textwidth}
\includegraphics[width=\textwidth]{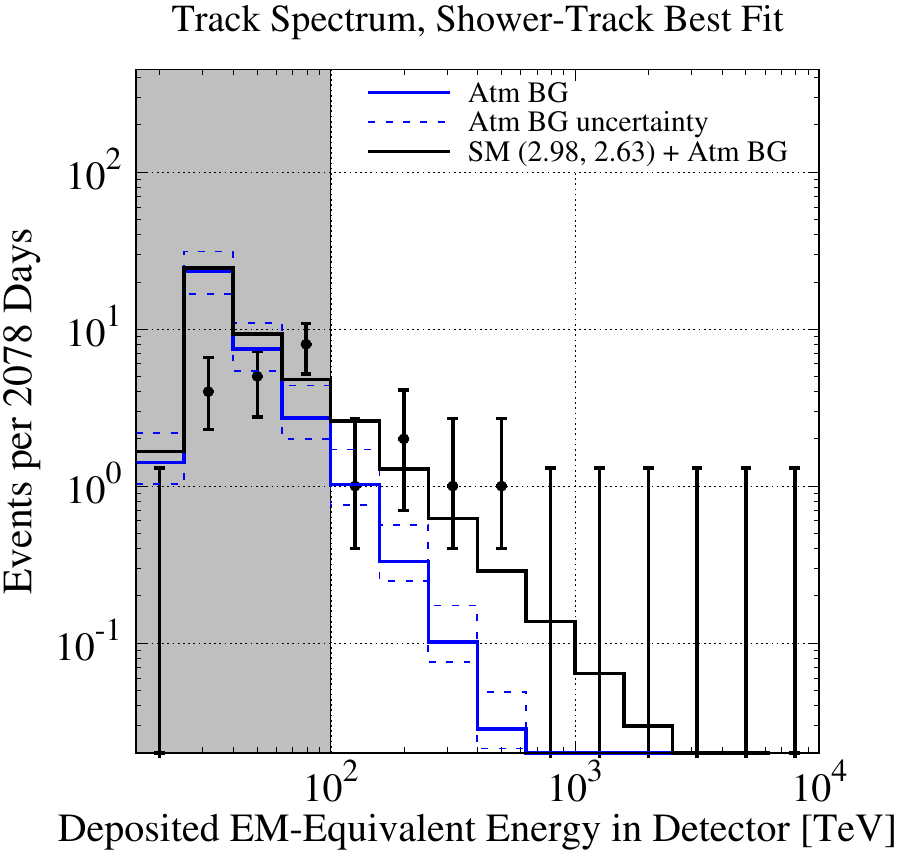}
\end{minipage}

\caption{Standard Model best fit of the IceCube 6-year dataset in the energy range [100 TeV, 10~PeV], which gives the values 
$\gamma=2.98$ and ${\rm C_0}= 2.63$ GeV cm$^{-2}$  str$^{-1}$  s$^{-1}$ for the neutrino flux parameters. The atmospheric background is computed from the estimated central values given in Ref.~\cite{Aartsen:2017mau}, from which we obtained $N_{\mu}=25.2$ and $N_{\text{atm}}=15.6$. The points in the grey region are not  included in the fit, as discussed in the text.}
\label{fig:shower-track-smbf}
\end{center}
\end{figure}

Assuming the flavor composition to be the standard one expected from neutrino 
oscillations, {\em i.e.}, $(f^\oplus_{\nu_e}:f^\oplus_{\nu_\mu}:f^\oplus_{\nu_\tau})=
(f^\oplus_{\bar \nu_e}:f^\oplus_{\bar\nu_\mu}:f^\oplus_{\bar\nu_\tau})
=(1:1:1)$, IceCube best fit of 6-year sample yields $\gamma=2.92^{+0.33}_{-0.29}$ and ${\rm C_0}= 2.46\pm 0.8$
GeV cm$^{-2}$  str$^{-1}$ s$^{-1}$~\cite{Aartsen:2017mau}. 
To test our fitting procedure we used the publicly available IceCube data, fit them to the SM and obtained~\footnote{The details about our
  statistical analysis considering the SM and several LQ
  scenarios are explained in Section~\ref{sec:4}. We advance the
  results from the best fit of the SM in this section both to
  visualize graphically the IceCube 6-year dataset,
  Fig.~{\ref{fig:shower-track-smbf}}, and also to establish some
  general but important remarks about the analysis.} 
\begin{align}
\gamma~=~2.98^{+0.42}_{-0.38}\,,\qquad\qquad {\rm C_0} =
2.63^{+1.87}_{-1.23}~\mathrm{GeV}\,\mathrm{cm}^{-2}\,\mathrm{str}^{-1}\,\mathrm{s}^{-1}
\end{align}
with a p-value of 0.25, in full agreement with the above-mentioned IceCube results~\cite{Aartsen:2017mau}. The best fit
curves for shower and track spectra are shown in
Fig.~{\ref{fig:shower-track-smbf}}.~\footnote{Notice that the contribution of the atmospheric background to the neutrino spectrum, which we obtain by using the modelling of~Ref.~\cite{Palomares-Ruiz:2015mka}, is consistent with the results shown in~Ref.~\cite{Aartsen:2017mau} concerning absolute numbers and spectrum shape. However, the validity of the model is challenged at low energies, for instance by the discrepancy in the number of tracks around 30 TeV, which can be treated in a way discussed in Refs.~\cite{Vincent:2016nut,Dey:2017ede}. Note that in this work we focus on the events above 100 TeV and rely on the validity of the atmospheric background model at high energies.}
For simplicity, in the fit we have used a fixed atmospheric background
spectrum which is derived from the central values for the total amount
of atmospheric neutrinos estimated in~\cite{Aartsen:2017mau}, which
are $N_{\mu}=25.2 \pm 7.3$ and
$N_{\text{atm}}=15.6^{+11.4}_{-03.9}$. To reduce the effects
of uncertainties we focus on the energy range [100 TeV, 10 PeV], in which the background contribution is sub-leading. Indeed,
the computation of the best fit considering the minimum (maximum)
expectations for the atmospheric background yields $\gamma =
3.0\,(2.9)$ and $C_0 = 2.7\,(2.4)$ with a p-value of 0.24 (0.23),
which lies perfectly inside the allowed 1$\sigma$ region. We will apply the same strategy to analyze LQ
scenarios. Before passing onto that discussion we examine some aspects of the simulation of IceCube events, with special attention to 
the new contributions introduced by LQ interactions.

\subsection{Calculation of the Event Rate in IceCube}

Icecube detects individual neutrino events through the Cherenkov
radiation emitted by charged secondary particles produced in the neutrino
interaction with the Antarctic ice.  These events are characterized by
the electromagnetic (EM) equivalent of the total deposited energy in the detector ($E_{\rm dep}$) 
and their topology (track or shower). Within the SM, track
events can only arise when $\nu_\mu$ or $\bar{\nu}_\mu$ interact with nucleons via charged current (CC) interactions, producing a energetic muon or antimuon.  On the other
hand, shower events can be induced by CC interactions of $\overset{\textbf{\fontsize{2pt}{2pt}\selectfont(---)}}{\nu_e}$ and $\overset{\textbf{\fontsize{2pt}{2pt}\selectfont(---)}}{\nu_\tau}$ or by neutral current (NC) interactions of all (anti-)neutrino flavors with nucleons.

To calculate the number of shower (sh) or track (tr) events
produced by SM interactions in an interval of reconstructed energy [$E_{\rm dep}$,$E_{\rm dep}+dE_{\rm
    dep}$], we first calculate the individual contributions to the event rate via the expression,
\bea
\frac{dN^{\rm c,h,X}_{\alpha}}{dE_{\rm dep}}= T N_A \int_{0}^\infty
dE_\nu \, {\cal A}^{\rm h}_{\alpha}(E_\nu) \frac{d\Phi_\alpha}{dE_\nu}
\int_{0}^{1} dy \, {\cal M}_{\rm eff}(E_{\rm t})\,R(E_{\rm t},E_{\rm
  dep},\sigma_{E_{\rm t}})\, \frac{d\sigma^{\rm X}_{\alpha}}{dy}\,,
\label{sm:rates}
\eea 
where $\rm c= sh,tr$ indicates the type of topology, $\rm h={S,N}$
the events entering the detector from the south (S) or north (N) hemisphere, and $\rm
X=\mathit{NC},\mathit{CC}, e$\textit{-scattering} stands for the different type of processes we consider. Here, $T$ is the exposure
time in seconds, $N_A~=~6.022~\times~10^{23}$~g$^{-1}$ is the Avogadro's
number times the number of moles per each gram of protons/neutrons, and ${\cal A}^{\rm h}_\alpha$ accounts for the effect of Earth's
attenuation, c.f.~Eq.\eqref{sec:att}. The effective mass of the detector in grams is denoted by ${\cal M}_{\rm eff}(E_{\rm t})$, which corresponds to the mass of
the target material convoluted with the efficiency of converting an event with 
true deposited energy $E_{\rm t}$ into the observed signal.
The resolution function, $R(E_{\rm t},E_{\rm dep},\sigma_{E_{\rm
  t}})$, is taken to be a Gaussian distribution with mean equal
to the true deposited energy $E_{\rm t}$ and standard deviation
$\sigma_{E_{\rm t}}$, both clearly dependent on the particular process
taking place.  Finally, $d\sigma^{\rm X}_{\alpha}/dy$ is the
differential cross section for process $\rm X$ in the inelasticity
interval $[y,y+dy]$. Our calculations were performed in the same way as described in
Refs.~\cite{Nunokawa:2016pop} and \cite{Palomares-Ruiz:2015mka}, and
we refer to these papers for further details.

\subsection{Leptoquark Contribution to the Event Rate} 

In our model, only the coupling $g_L^{c\tau}$ affects the neutrino-nucleon production
cross section at IceCube, c.f.~Eq.~\eqref{eq:slq1}. The other coupling, $g_R^{b\tau}$, only
contributes to the total decay width of the LQ bosons. In this case, since the new interactions only involves quarks, neutrinos and  
$\tau$-leptons, this scenario will only produce nonstandard shower-like events in IceCube. The event rate with $\nu_\tau$ in Eq.~\eqref{sm:rates} will then receive the New Physics contribution

\bea 
\frac{dN^{\rm sh,h,\Delta}_{\nu_\tau}}{dE_{\rm dep}}= T N_A \int_{0}^\infty
dE_\nu \, {\cal A}^{\rm h}_{\nu_\tau}(E_\nu)
\frac{d\Phi_{\nu_\tau}}{dE_\nu} \int_{y_{\rm min}}^{y_{\rm max}} dy \,
     {\cal M}_{\rm eff}(E_t)\,R(E_{t},E_{\rm
       dep},\sigma_{E_t})\, \frac{d\sigma^{\rm
         \Delta}_{\nu_\tau}}{dy},\,
\label{lq:rate}
\eea

\noindent  where the differential cross-section expressions $d\sigma_{\nu_\tau}^\Delta/dy$ can be found in Appendix~\ref{lqxsec}, with $y$ being the inelasticiy variable. The same expression applies \textit{mutatis mutandis} to the  $\bar{\nu}_\tau$ event rate. 

\begin{figure}[htb]
\centering
\includegraphics[width=0.7\linewidth]{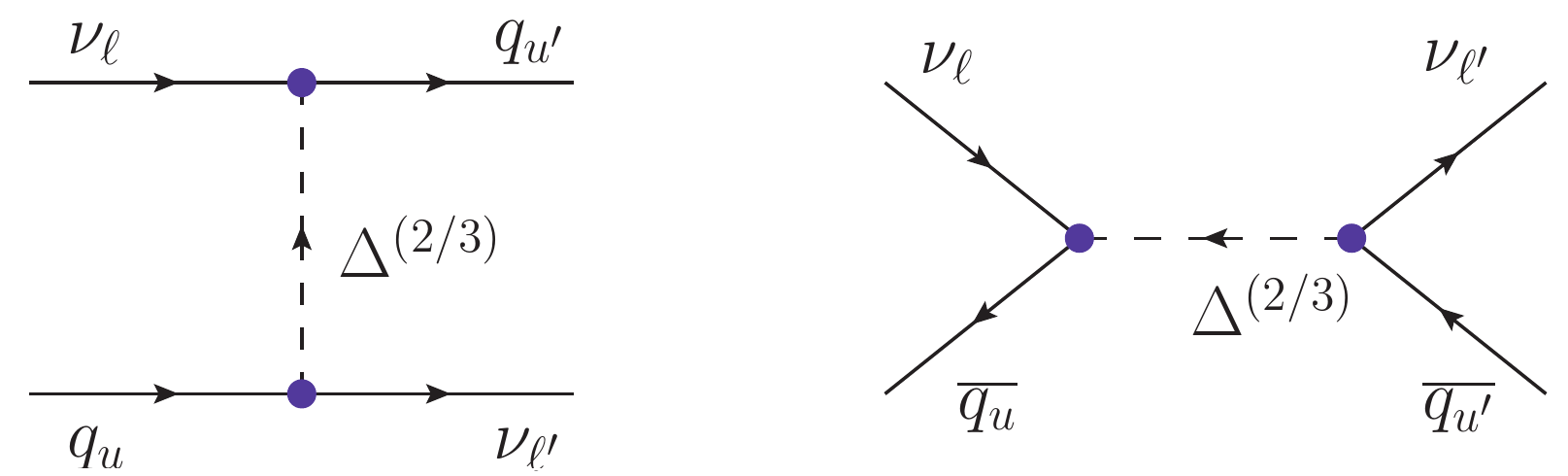}
\caption{Feynman diagrams contributing to $\nu_\ell + q_u \to {\nu}_{\ell^\prime} + q_{u^\prime}$ and $\nu_\ell + \bar{q}_u \to {\nu}_{\ell^\prime} + \bar{q}_{u^\prime}$ via a LQ exchange.
Here $\nu_\ell, \nu_{\ell'}$ are generic neutrino flavors and $q_u, q_{u'}$ are generic up-type quarks.}
\label{fig:nuq}
\end{figure}

\begin{figure}[htb!]
\centering
\includegraphics[width=0.7\linewidth]{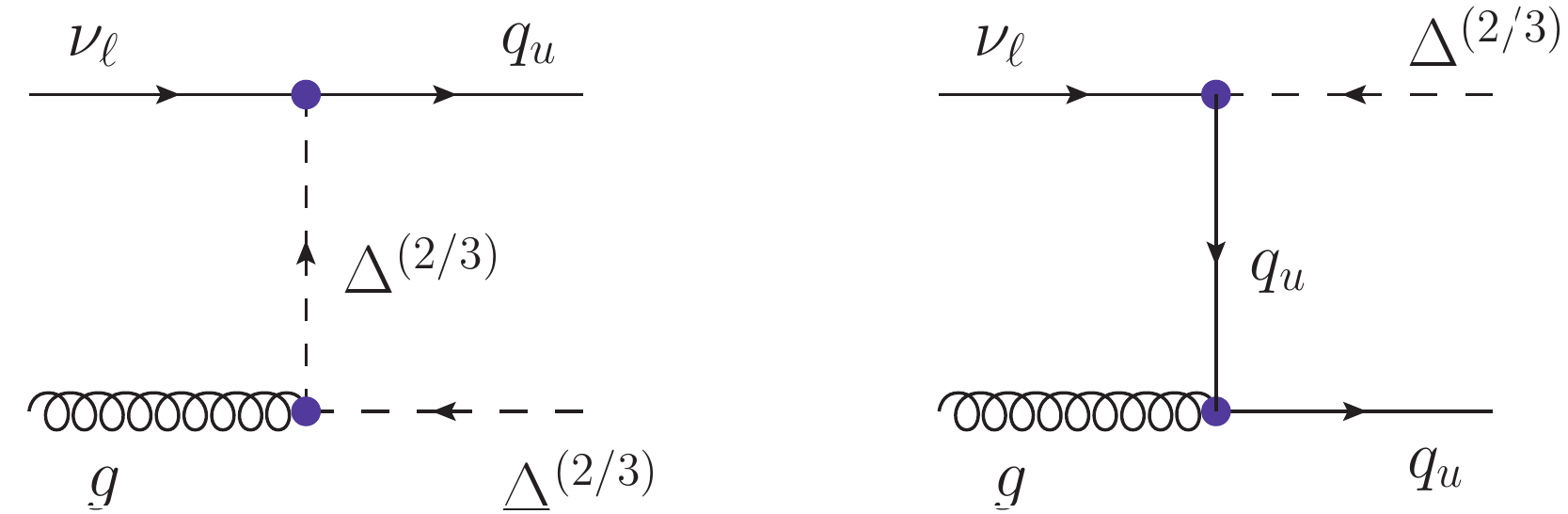}
\caption{Feynman diagrams contributing to the process $g + \nu_\ell \to q_u + \Delta$, where the LQ particle decays into $\bar{q_u}+\nu_{\ell^\prime}$. }
\label{fig:gluon}
\end{figure}

\begin{figure}[htb!]
\begin{center}
\begin{minipage}[b]{0.48\textwidth}
\includegraphics[width=\textwidth]{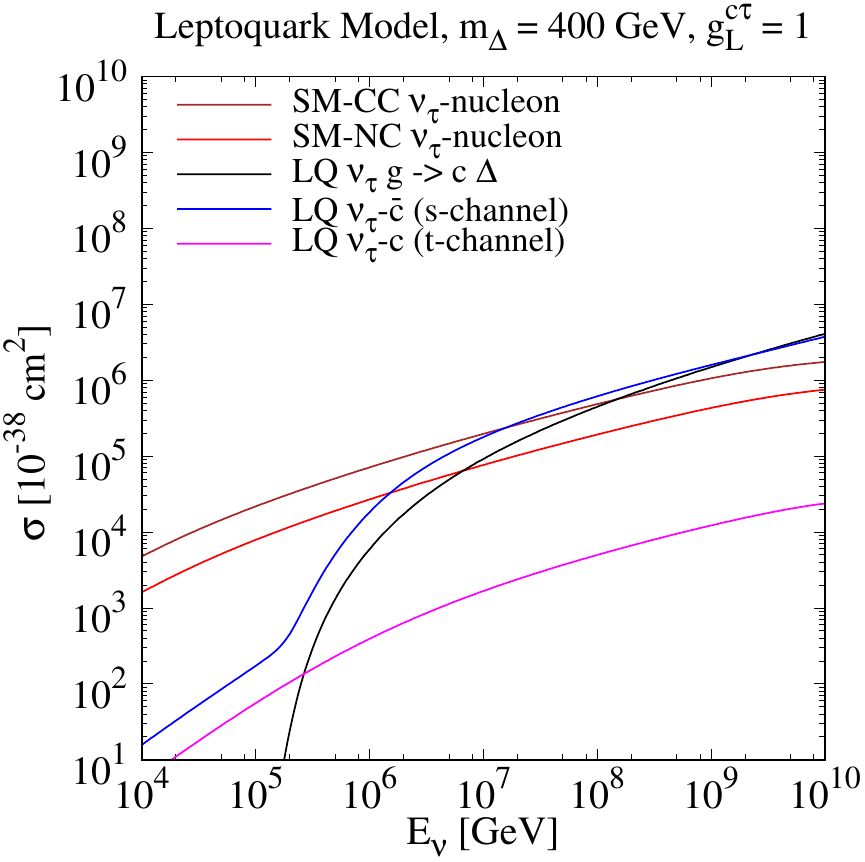} 
\end{minipage}
\hfill
\begin{minipage}[b]{0.48\textwidth}
\includegraphics[width=\textwidth]{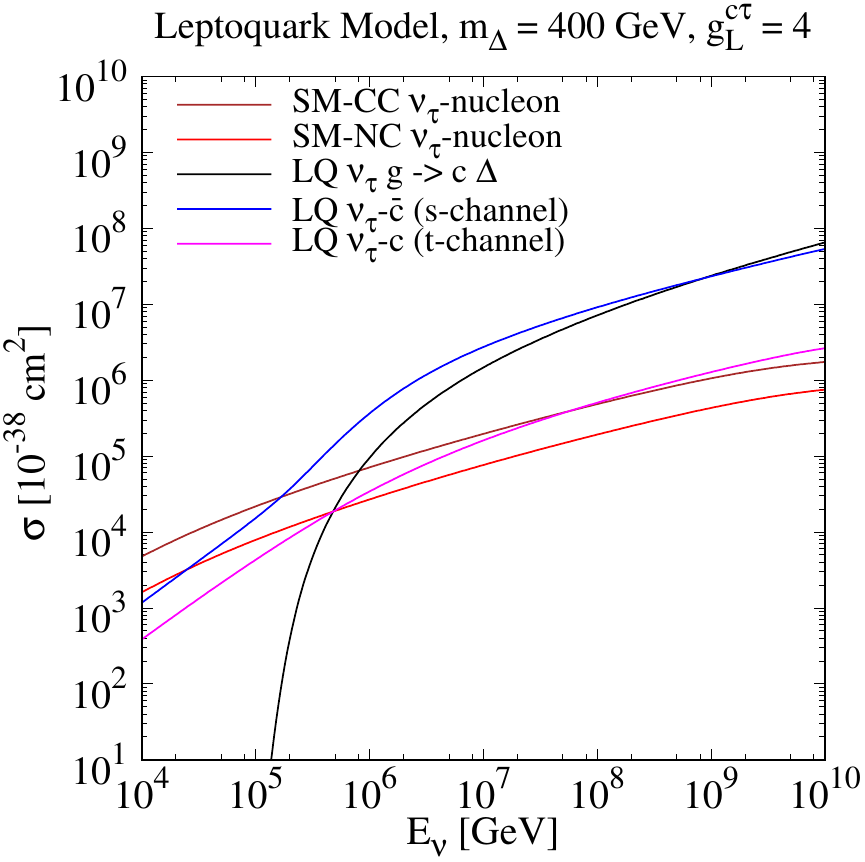}
\end{minipage}

\caption{$\nu_\tau$-nucleon cross sections: 
CC contribution in the SM,
NC contribution in the SM, 
$\nu_\tau$-gluon contribution,
$\nu_\tau$-c-quark s-channel contribution,
$\nu_\tau$-c-quark t-channel contribution.
For the LQ cross sections $m_\Delta=400$ GeV and 
$g_L^{c\tau}=1$ (left) and $g_L^{c\tau}=4$ (right). }
\label{x-sections}
\end{center}
\end{figure}

In previous LQ studies~\cite{Anchordoqui:2006wc,Barger:2013pla,Dutta:2015dka,Dey:2017ede,Mileo:2016zeo,Dey:2015eaa,Chauhan:2017ndd}, only the parton level processes $\nu_\ell + \overset{\textbf{\fontsize{2pt}{2pt}\selectfont(---)}}{q} \to \nu_\ell + \overset{\textbf{\fontsize{2pt}{2pt}\selectfont(---)}}{q}$ were considered in the cross-section, c.f. Fig.~\ref{fig:nuq}. In addition to the interaction with
quarks, high-energy neutrinos can also interact with gluons via
the transition $\nu_\ell +g \to q_u+\Delta$, with the LQ decaying inside the
detector, as shown in
Fig.~\ref{fig:gluon}. This contribution to the cross-section can be of the same order of the $\nu_\ell + q$ contribution when dealing with heavy quarks, $q\in \lbrace c,b\rbrace$, and thus cannot be neglected, as we illustrate in Fig.~\ref{x-sections}. This can be understood from the suppression of the $s$- and $t$-channel interactions with heavy quarks by the small values of the parton distribution functions (PDFs). Furthermore, the gluon PDF shows a fast growth when the Bjorken scale variable $x$ approaches zero, giving a non-negligible contribution to heavy quark cross-sections.

\begin{figure}[htb]
\centering

\begin{minipage}[b]{0.48\textwidth}
\includegraphics[width=\textwidth]{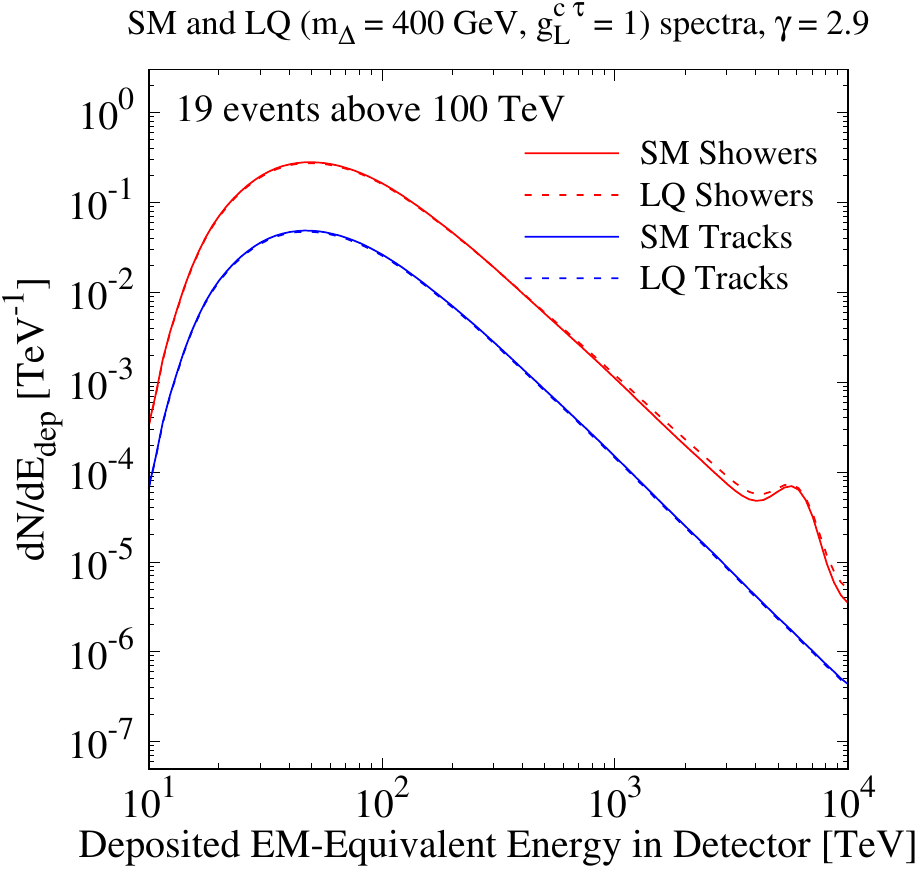}
\end{minipage}
\hfill
\begin{minipage}[b]{0.48\textwidth}
\includegraphics[width=\textwidth]{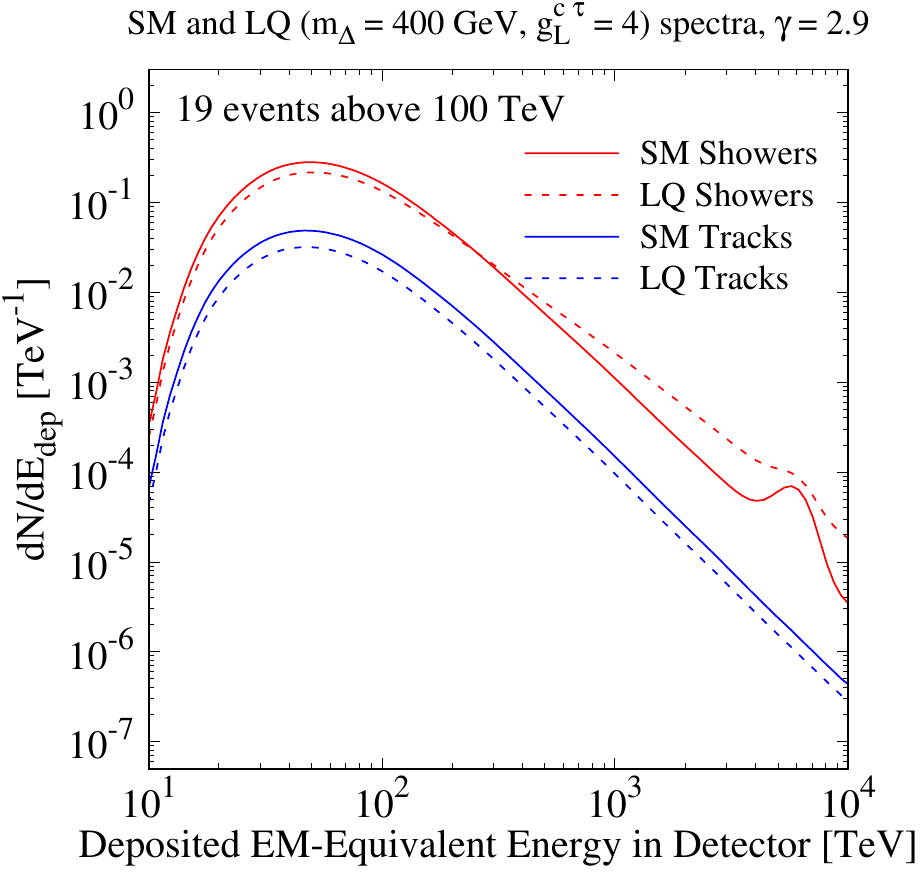}
\end{minipage}

\caption{Comparison between the distribution of shower and track
  events for the SM and two representative LQ scenarios:
  $m_\Delta= 400$ GeV and $g_L^{c\tau} = 1$ (left panel) and $m_\Delta=
  400$ GeV and $g_L^{c\tau} = 4$ (right panel). Notice that the LQ shower
  (and track) distributions include both the contribution of the SM as well as of the  LQ interactions.}
\label{dNdE}
\end{figure}

To estimate the impact of LQ interactions to the IceCube data we compare  in Fig.~\ref{dNdE} the distribution of shower and track events generated by the SM and two representative LQ scenarios. These results were obtained under the restriction of a fixed total number of events. We see that the nonstandard effects appear only at high neutrino energies ($E_\nu\gtrsim 500$~TeV), anticipating a modest sensitivity to LQs with current IceCube
data. 
This instigates us to investigate the situation with maximal 
future sensitivity of IceCube and IceCube-Gen2~\cite{Aartsen:2015dkp} which we postpone to Sec.~\ref{sec:4}.

\subsection{Attenuation and Regeneration Effects} 
\label{sec:att}

The neutrinos that arrive in IceCube from the south hemisphere do not suffer 
Earth effects, having then an attenuation factor of ${\cal A}^{\rm S}_{\alpha}(E_\nu)=\frac{1}{2}$ for an isotropic flux in Eq.~\eqref{sm:rates}.
The neutrinos entering from the north hemisphere have to go through the Earth before 
reaching the IceCube detector with the attenuation factor of the resulting flux given by
\begin{equation}
{\cal A}^{\rm N}_{\alpha}(E_\nu) = \frac{1}{2} \int_0^{1} {\cal A}_{\alpha}(E_\nu,\theta) 
\;\; d \cos \theta \, ,
\end{equation}
where $\theta$ is the nadir angle, and
\begin{equation}
{\cal A}_{\alpha}(E_\nu,\theta) \equiv \frac{{\cal F}_{\alpha}(E_\nu,X(\theta))}{
{\cal F}_{\alpha}(E_\nu,0)}\, ,
\end{equation}
is simply a fraction of the
initial flux after propagating though a column depth 
$X(\theta)$,
\bea
X(\theta) = \int_0^{L(\theta)} \rho(s) \, ds \, ,
\eea
where $\rho(s)$ is the matter density at the point $s$,  
$L(\theta)= 2 R_{\oplus} \cos \theta$, and $R_\oplus$ is the Earth's radius. 
In the above formulas, the function ${\cal F}_{\alpha}$ is defined by
\begin{equation}
{\cal F}_{\alpha}(E,X(\theta)) \equiv \frac{\partial\Phi_\alpha}{\partial E}(E,X(\theta))
\, .
\end{equation}
To calculate ${\cal A}^{\rm N}_{\alpha}(E_\nu)$ we need to solve the coupled integro-differential transport equations given in Appendix~\ref{transport} which include absorption and regeneration effects via particle decays. We have followed the prescription described in Ref.~\cite{Rakshit:2006yi} 
to numerically solve these equations. Furthermore, we assumed that the absorption of $\tau$ and $\Delta^{(2/3)}$ is only given 
by their decays into lighter particles, neglecting the subdominant scattering contributions.

\begin{figure}[htb]
\centering

\begin{minipage}[b]{0.48\textwidth}
\includegraphics[width=\textwidth]{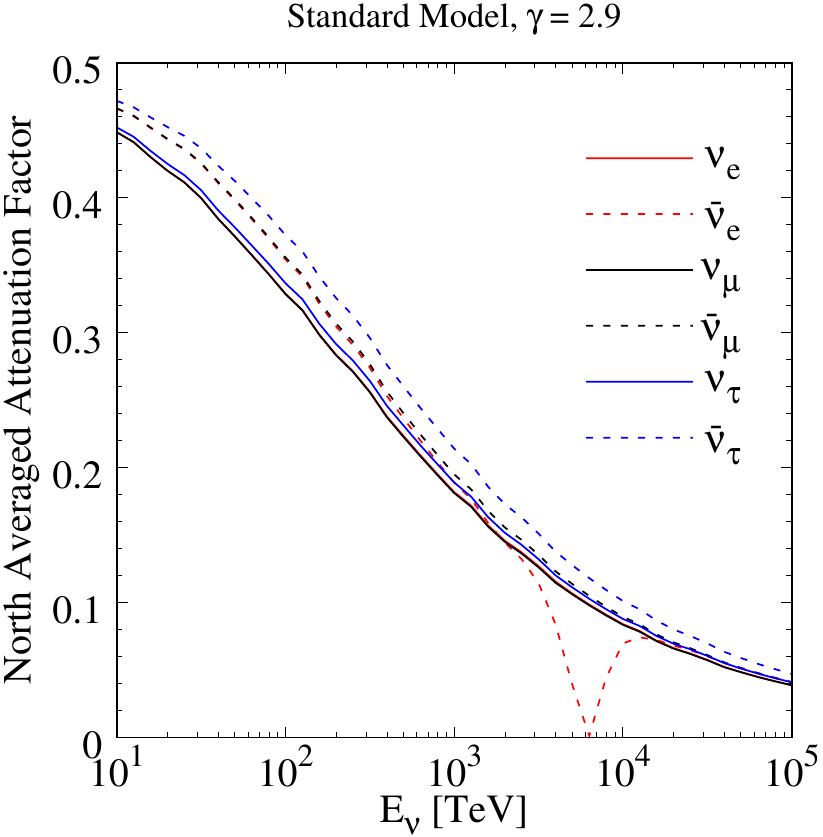}
\end{minipage}
\hfill
\begin{minipage}[b]{0.48\textwidth}
\includegraphics[width=\textwidth]{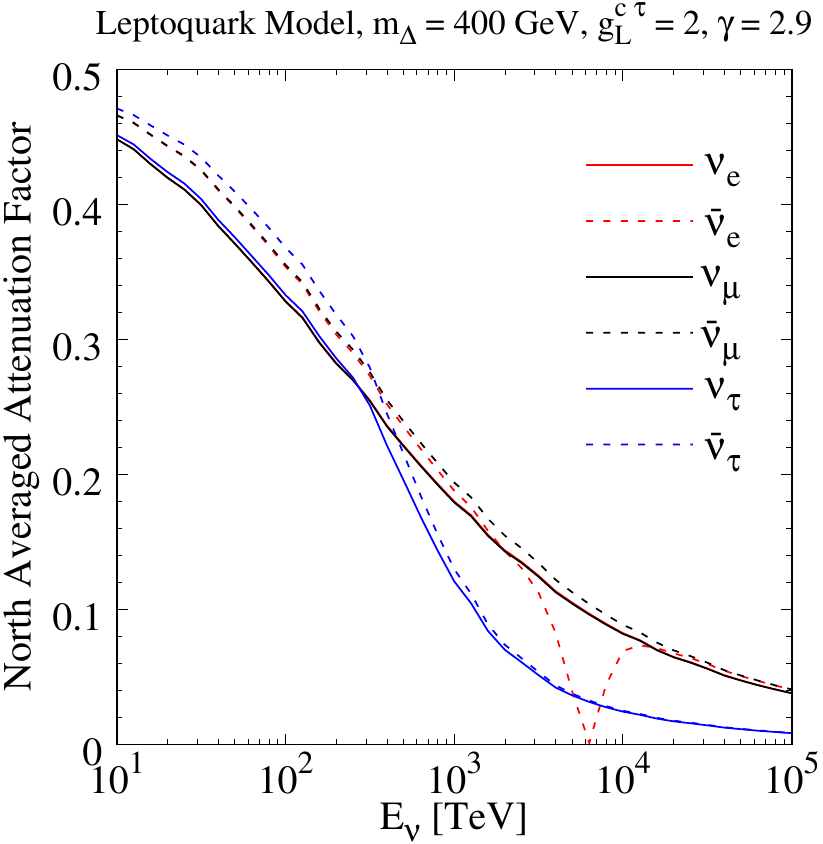}
\end{minipage}

\caption{Average attenuation factor ${\cal A}^{\rm N}_{\alpha}(E_\nu)$ 
for each (anti-)neutrino flavor for the SM (left panel) and for a LQ model with $m_\Delta= 400$ GeV and $g_L^{c\tau} = 2$ (right panel).
These results has been obtained by fixing $\gamma=2.9$, but they only show a mild dependence on this parameter.}
\label{attenuation}
\end{figure}

The average attenuation factor ${\cal A}^{\rm N}_{\alpha}(E_\nu)$
for each (anti-)neutrino flavor is shown in Fig.~\ref{attenuation} for the SM (left panel), and for a LQ scenario with $m_\Delta= 400$ GeV and $g_L^{c\tau}=2$ (righ panel).
As expected, the attenuation factors ${\cal A}^{\rm N}_{\nu_\tau}$ (${\cal A}^{\rm N}_{\bar \nu_\tau}$) are highly 
affected by LQ interactions for $E_\nu \gtrsim 500$ TeV. Note, however, that since the incoming flux falls rapidly with $E_\nu$, this effect turns out to be smaller than suggested by Fig.~\ref{attenuation}. Furthermore, we included the regeneration effects in our simulation which can lower the attenuation factors by at most 10\%.

\section{\label{sec:4} Statistical Data Analysis at IceCube, Present and Future}

As stated in previous Sections, we use the 6-year HESE sample in the energy range [100 TeV, 10 PeV]~\cite{Aartsen:2017mau}. To quantify
the goodness-of-fit of each model we follow the maximum likelihood
method described in Ref.~\cite{Aartsen:2015knd}. The test statistic is defined by

\begin{equation}
-2\ln L(H)/L_{\text{sat}}=-2\sum_{i}^{N_{\text{bins}}}\ln(e^{o_{i}-n_{i}}(n_{i}/o_{i})^{o_{i}})\,,
\label{eq:ts1}
\end{equation}

\noindent where $L(H)$ is the likelihood of the tested hypothesis $H$ (product of Poisson
probabilities), and $L_{\text{sat}}$ is
the likelihood of the model that predicts the exact observed outcome,
$\mbox{i.e.}\,\,n_{i}\equiv o_{i}$. Moreover, the values of
$o_{i}$ and $n_{i}$ correspond to the observed and predicted number of events in the bin $i$, respectively. We divide the events sample in the neutrino energy range
[100 TeV, 10 PeV] in $10$ logarithmic energy bins for tracks and as many for showers. The total number of bins, $N_{\text{bins}}=20$, is comparable to the 28 events detected above 100 TeV. We consider a track to shower misidentification factor
(missID) of 20\%, which is reasonable considering that IceCube has
measured it to be approximately 30\% for low energy
tracks~\cite{Aartsen:2015ivb} and since it is expected to decrease with
energy. Nevertheless, we have verified that the final results are unaffected by using a 10\% or a 30\% missID factor, similarly to what has been obtained in Ref.~\cite{Dey:2017ede}. 

We reiterate that in our statistical analysis we consider 5 parameters:
\begin{equation}
\gamma ,\,  C_0, \, m_\Delta, \, g_L^{c\tau},\, g_R^{b\tau}, 
\end{equation}
which we vary as $m_\Delta \in [300\,\text{GeV},700\,\text{GeV}]$,  $\vert g_L^{c\tau}\vert\in [1,4]$, and we set $g_R^{b\tau}\approx 0$ since it does not directly contribute to the cross-sections relevant to this study.

\subsection{Current Data Analysis}

To evaluate the quality of the fit of a hypothesis H with
respect to current data, we compute the minimum of the test statistic~\eqref{eq:ts1} and its corresponding
p-value, obtained by simulating pseudo-data with the best fit values of $\gamma$ and $C_0$. As already mentioned in the text, we find that the SM best fit point yields $\gamma=2.98$ and ${\rm C_0}= 2.63$ GeV cm$^{-2}$ str$^{-1}$ s$^{-1}$ with a test statistic value of 17.05, and a p-value of 0.25. This indicates that the SM compatibility with the current data is quite acceptable. 

\begin{figure}[htb]
\begin{minipage}[b]{0.50\textwidth}
\includegraphics[width=\textwidth]{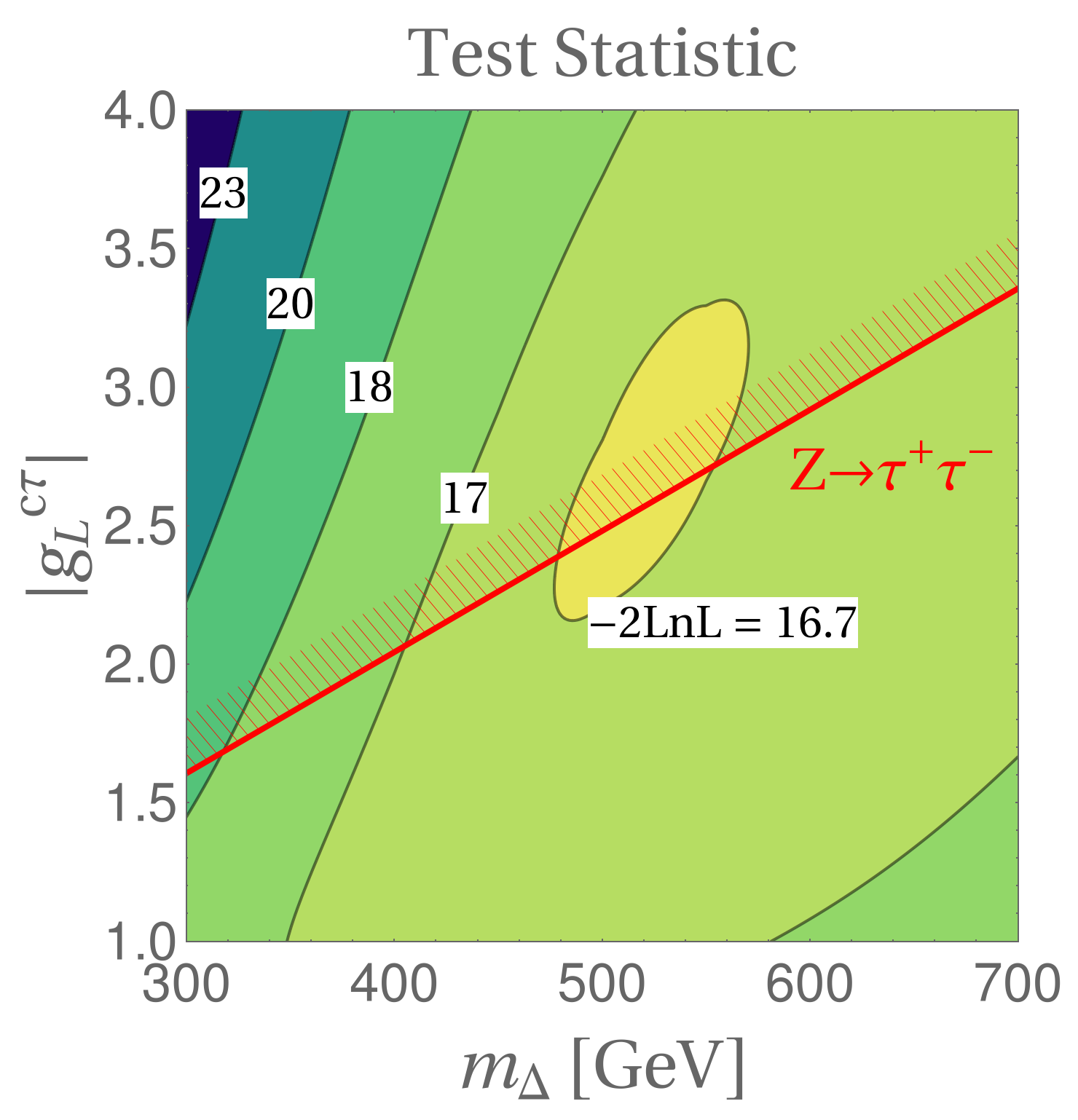}
\end{minipage}
\hfill
\begin{minipage}[b]{0.50\textwidth}
\includegraphics[width=\textwidth]{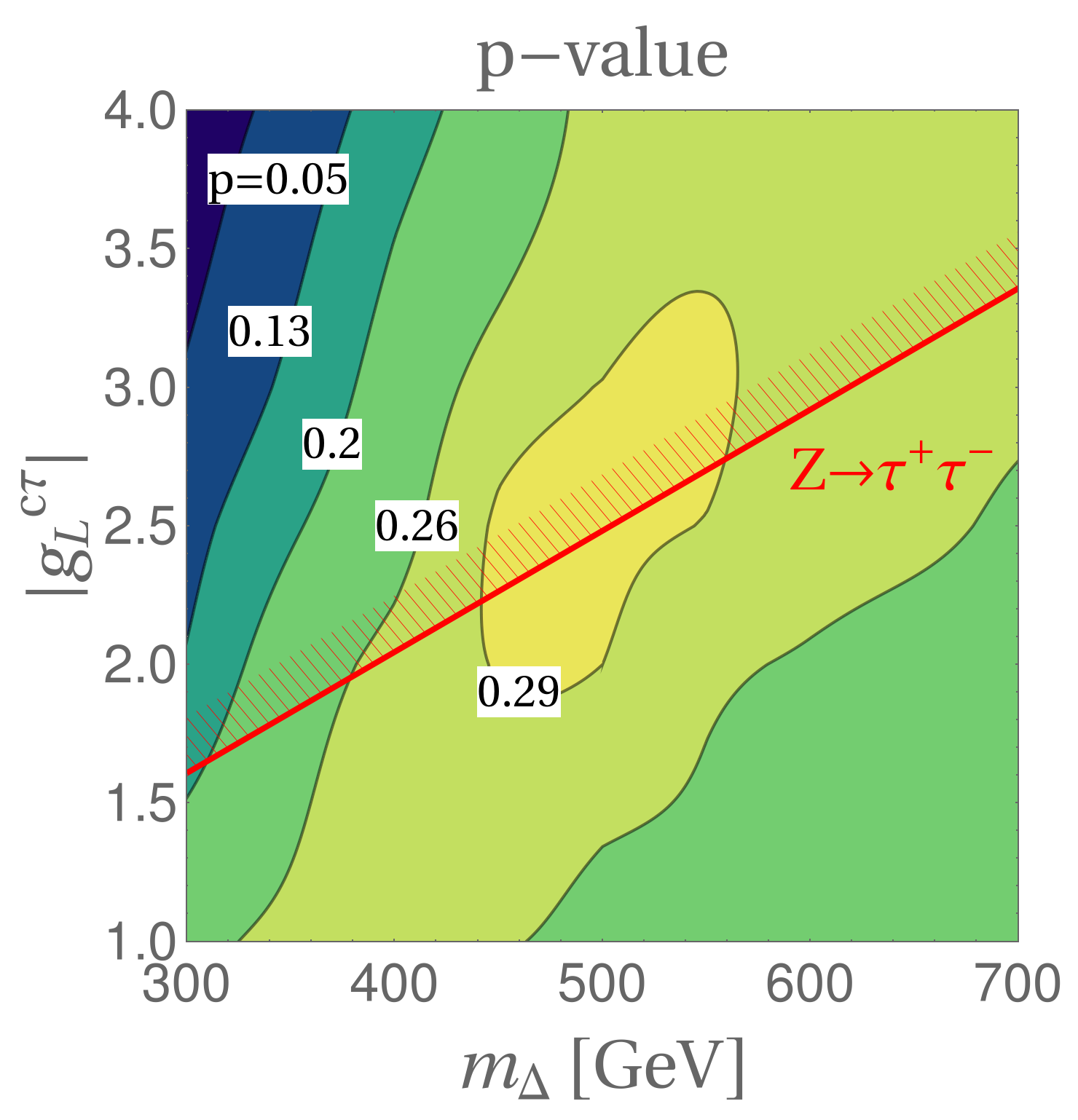}
\end{minipage}
\caption{Summary of the best fit analysis of LQ models characterized by the values of $m_{\Delta}$ and $\vert g_{c\tau}\vert$.
 These results are based on the minimization of the test statistic defined in
  Eq.~(\ref{eq:ts1}) with respect to $\gamma$ and $C_0$ using the IceCube
  6-year data sample.  In the left panel we show contour lines for the
  values of the test statistic, while in the right panel we also plot
  contour lines but considering the obtained p-value.  We show in both
  figures the exclusion coming from the $Z$-pole observables discussed
  in Sec.~\ref{sec:cons}.}
\label{fig:StatisticalTestCurrentData}
\end{figure}

\begin{figure}[htb]
\begin{center}
\begin{minipage}[b]{0.48\textwidth}
\includegraphics[width=\textwidth]{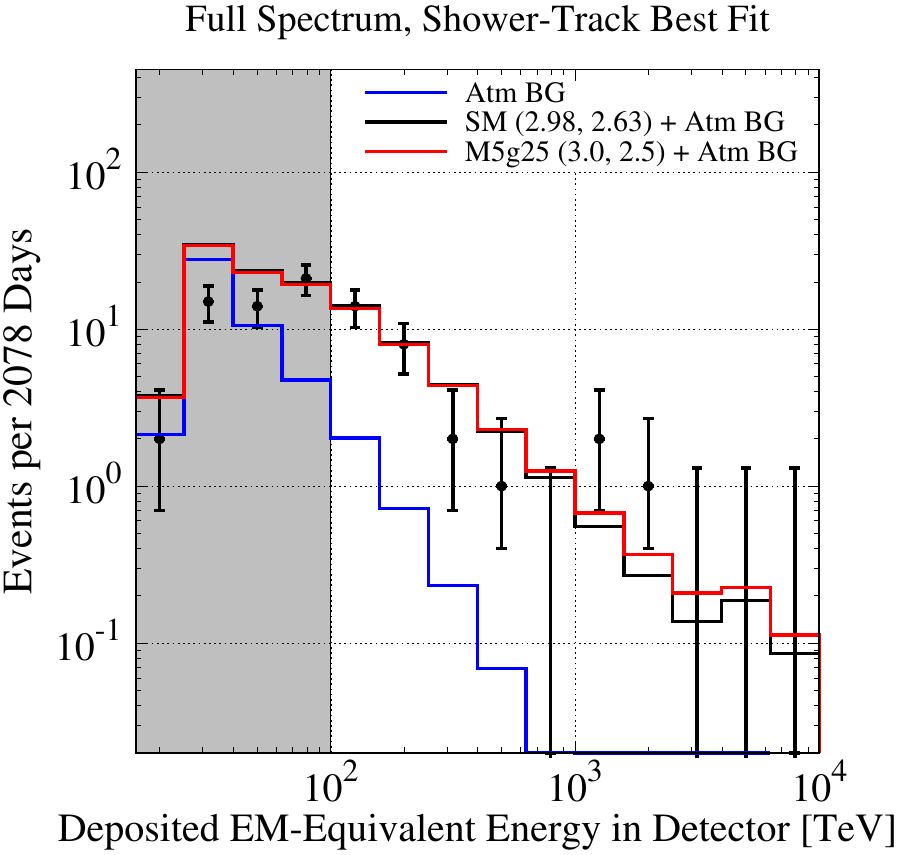}
\end{minipage}
\hfill
\begin{minipage}[b]{0.48\textwidth}
\includegraphics[width=\textwidth]{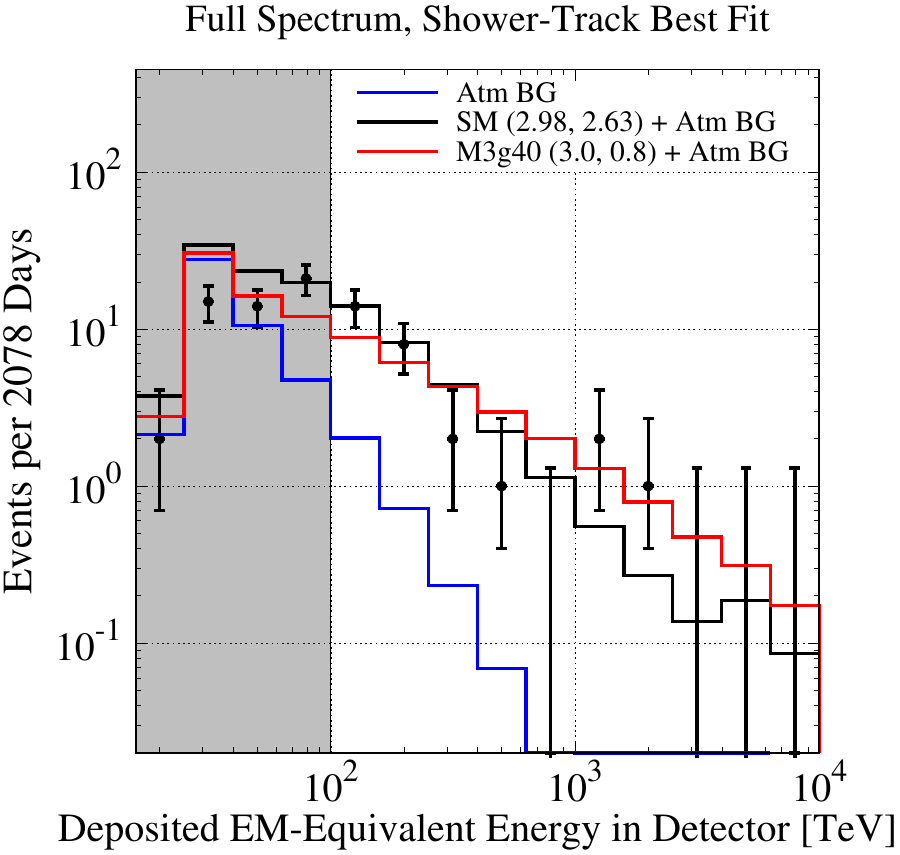}
\end{minipage}
\caption{Comparison between the SM and LQ best fit curves to
  the IceCube 6-year dataset. In the left panel we show the best fit curve
  for $(m_\Delta,\,\vert g_L^{c\tau}\vert) = (500\,\text{GeV},\,2.5)$, which is
  one representative case of scenarios with $-2\ln L = 16.7$. In the
  right panel, we show the best fit curve for
  $(m_\Delta,\,\vert g_L^{c\tau}\vert) = (300\,\text{GeV},\,4)$, which has a
  p-value around 0.05 and therefore it is in tension with current
  data. In the plot legend we quote within parentheses the best fit
 values of $\gamma$ and $C_0$ for each model.}
\label{fig:SMLQbestFit}
\end{center}
\end{figure}

The results obtained in the LQ scenario are shown in Fig.~\ref{fig:StatisticalTestCurrentData} for different values of LQ mass $m_\Delta$ and coupling $|g_L^{c\tau}|$. We find that the best fit point is given by $(m_\Delta,\,\vert g_L^{c\tau}\vert) = (500\,\text{GeV},\,2.5)$ with a minimum test statistic of 16.7, and therefore the event distribution of the SM and the best LQ
model are indistinguishable from the IceCube data, also shown in left panel of Fig.~\ref{fig:SMLQbestFit}. From the right panel of Fig.~\ref{fig:SMLQbestFit} we learn that only LQ models with very large $g_L^{c\tau} \approx 3.5$ and small masses $m_\Delta \approx 300$~GeV can be excluded by current IceCube data to $95\%$ CL ($p\leq 0.05$). These parameters, however, are already excluded by the limits arising from the $Z$-pole observables, as discussed in Sec.~\ref{sec:1}. 

\subsection{Future IceCube(-Gen2) Projections}

To investigate the projected sensitivity of IceCube to our LQ scenario, we compute the projected exclusion regions in the plane $(m_\Delta,\,\vert g_L^{c\tau}\vert)$ by assuming the SM is the null hypothesis. We considered the test statistic defined
by the likelihood ratio between the LQ and the SM hypotheses,
given by $q=-2\log L(\text{LQ})/L(\text{SM})$ with $L(\text{LQ})$ and
$L(\text{SM})$ defined analogously to Eq.~(\ref{eq:ts1}). The value of
$q$ is obtained after the minimization of each hypothesis with respect
to $(\gamma, C_0)$ by using projected data, which is generated from Monte Carlo simulations of the SM best fit to current data.
More specifically, we evaluate the luminosity $\mathcal{L}$, at which the projected data is simulated, by using multiples of current 6-year exposure. To establish a
criteria for exclusion we consider the p-value associated to each 
tested value of $q$. We simply assume that $q$ follows a
$\chi^{2}$-distribution with $k$ degrees of freedom, where $k$ is the
difference in the number of parameters between the LQ and SM
hypotheses, which in our case is $k=2$.  Thus, for a given value of
$q$, the p-value is given by $p=\exp(-q/2)$.

\begin{figure}[htb]
\begin{center}
\includegraphics[width=0.65\textwidth]{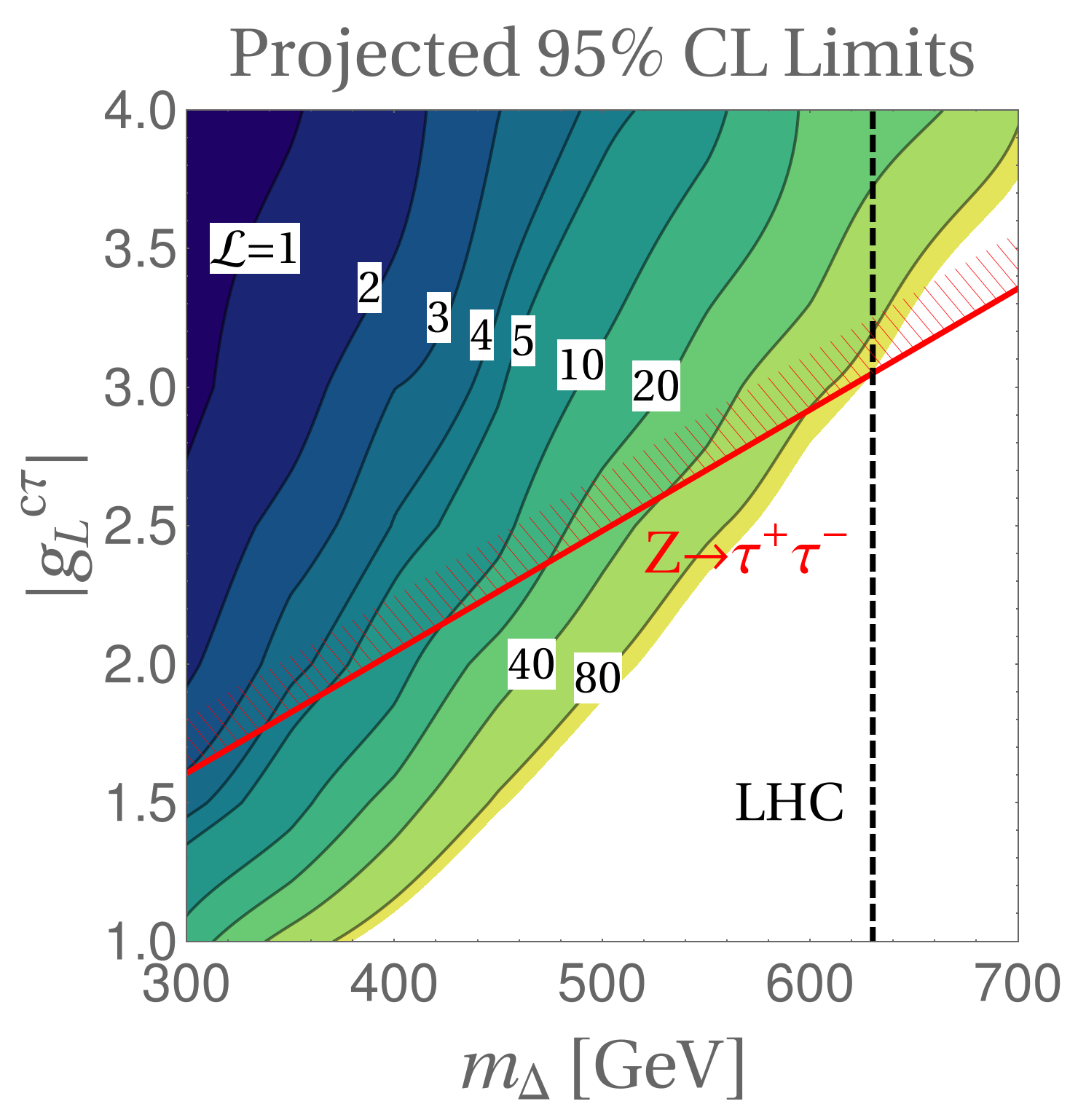}
\caption{Estimation of the 95\% CL exclusion regions in the parameter space ($m_{\Delta}$, $\vert g_{L}^{c\tau}\vert$) 
considering several projections of IceCube data at exposures measured in units of the current 6-year data sample. 
We also show the constraints on large LQ couplings derived from the $Z$-pole observables (red line) and the lower limits on the LQ mass stemming from the reinterpretation of LHC limits on pair-produced LQs (dashed line), see~Sec.~\ref{sec:cons} for details.}
\label{fig:Projections}
\end{center}
\end{figure}

In Fig.~\ref{fig:Projections} we plot our results as the projected 95\% CL exclusion regions in the $(m_\Delta,\,\vert g_L^{c\tau}\vert)$ plane, along with the limits arising from the $Z$-pole observables, as well as the region eliminated by the reinterpretation of LHC searches for pair-produced LQs decaying into $b\tau$~\cite{Khachatryan:2016jqo,CMS:2016hsa}, $c\nu$~\cite{CMS:2018bhq}  and $t\tau$~\cite{Khachatryan:2015bsa,Sirunyan:2018nkj}, c.f.~discussion in Section~\ref{sec:1}. We see that the parameter region accessible to IceCube increases at a modest pace with the luminosity $\mathcal{L}$, reaching the corners ($m_{\Delta}=300$, $\vert g_{L}^{c\tau}\vert=1$) and ($m_{\Delta}=600$, $\vert g_{L}^{c\tau}\vert =4$) only with about $20$ and $40$ times more events than the current data. An exposure equivalent to $10$ ($40$) times the current one might be within reach of IceCube-Gen2~\cite{Aartsen:2015dkp} 
after about $5$ ($20$)  years of data taking. From that plot we also see that the region of parameters accessible to IceCube(-Gen2) is already disfavored by a combination of constraints arising from the direct (LHC) and indirect searches (flavor physics and LEP).

\section{\label{sec:conc} Conclusions}

Several authors have inquired about the potential of IceCube detector in the South Pole
to search for LQs~\cite{Anchordoqui:2006wc,Barger:2013pla,Dutta:2015dka,Mileo:2016zeo,Dey:2015eaa,Chauhan:2017ndd,Dey:2017ede}. In most of these works LQs couple to the first
generation of quarks. On the one hand this allows LQ production to profit from the valence quark 
contribution to the neutrino-nucleon cross section. On the other hand limits from direct searches,
 atomic parity violation and flavor physics are at odds with these possibilities.
 
Here, instead, we focus on LQs that couple to heavy quark
flavors, a scenario that recently became of particular interest in connection
with $B$-meson semileptonic decays.  The hope that
IceCube could contribute to the search of this type of LQ models
springs from a few considerations. First, these models cannot be dismissed by current direct or indirect detection searches. Second, 
their production cross section has been underestimated in the literature
as the neutrino-gluon contribution has been neglected up to now.
Third, attenuation effects when crossing the Earth are modified 
by the new LQ interactions which was not  
considered previously. Finally, the proposal of IceCube-Gen2~\cite{Aartsen:2015dkp} 
allows for exploring a much higher statistics of HESE  than at 
the present time.

In this paper we focused on a simple LQ scenario that can explain $R_{D^{(\ast)}}^\mathrm{exp} > R_{D^{(\ast)}}^\mathrm{SM}$ 
and involves only three new parameters: the LQ mass ($m_\Delta$) and 
two couplings ($g_L^{c\tau}$ and $g_R^{b\tau}$). Parameter space is constrained in such a way as 
to solve $R_{D^{(\ast)}}$ and satisfy experimental constraints coming from the $Z$-decay modes. With parameters constrained in this way we estimate the present (future) 
sensitivity of IceCube (IceCube-Gen2) to this model. Our results are summarized in Fig.~\ref{fig:Projections} from which we see that the parameter space accessible to IceCube/IceCube-Gen2 
is already excluded by the direct searches (see 
Fig.~\ref{fig:direct-searches}) and the low-energy flavor physics observables.
That conclusion is based on our current 
understanding of the HESE data and the value of the spectral index $\gamma$ in particular. If that value turns out to be lower, then also Fig.~\ref{fig:Projections} would change because there would be more 
data corresponding to heavier LQs but with the mass still lower than about $650$~GeV, to comply with the results of the direct searches at LHC.

We should also stress that in our calculation we consistently used a democratic 
flavor distribution of the incoming flux, which is plausible and compatible with the current 
data. Other possibilities are, however, possible too which is one reason more why settling that issue is of major importance for
searching the effects of physics beyond the SM~\cite{Aartsen:2017mau}.

\begin{acknowledgments} 
This work was partially supported by Funda\c{c}\~ao de Amparo \`a
Pesquisa do Estado de S\~ao Paulo (FAPESP) and Conselho Nacional de
Ci\^encia e Tecnologia (CNPq).  R.Z.F.~is grateful for the hospitality
of the Laboratoire de Physique Th\'eorique of Universit\'e Paris-Sud
during the final part of this work and acknowledges the CNRS for
support during this time.  D.B.~acknow\-led\-ges the INP-CNRS
support via{\sl ``Inphyniti"}. This project has also received support
from the European Union's Horizon 2020 research and innovation
programme under the Marie Sklodowska-Curie grant agreement 
N$^\circ$ 690575 and 674896.
\end{acknowledgments}

\clearpage

\appendix

\section{Leptoquark Contributions to the Neutrino-Nucleon Scattering Cross Section}
\label{lqxsec}

For completeness we give here the expressions for the neutrino-nucleon scattering cross section at the parton level. The total differential cross section induced by LQs can be written as
\bea
\frac{d\sigma^{\Delta}_{\nu_{\tau}}}{dy}(y,E_\nu) &=& \int_{x_{\rm min}}^{1} dx \, g(x,\hat s) \, 
\frac{\partial^2\sigma^{g\nu_\tau}}{\partial x \partial y}(E_\nu,x,y)\nonumber \\
&+& \sum _{q}\int_{0}^{1} dx \, q(x,\hat s) \, 
\frac{\partial^2\sigma^{q\nu_\tau}}{\partial x \partial y}(E_\nu,x,y)\,,
\eea
for an incident neutrino $\nu_\tau$. Here 
$g(x,\hat s)$ ($q(x,\hat s)$) is the gluon (quark $q$) 
PDF of the nucleon. In our computaions, we considered the NNPDF2.3 PDF sets~\cite{Ball:2012cx}, which are implemented in LHAPDF6~\cite{Buckley:2014ana}.  

\subsection{$\nu_\ell$-gluon contribution}
We give here the expressions for the gluon-neutrino differential 
cross sections induced by LQs. The kinematics of these processes can be described by the following variables

$$y = \frac{E_\nu-E_\Delta}{E_\nu}\, , \quad \quad x = \frac{Q^2}{2
  m_N E_\nu \, y}\,, \quad \quad \hat s = 2 x \, m_N E_\nu \, , $$
with the inelasticity $y$ and the Bjorken variable $x$ defined in the intervals~\footnote{We define $t=(p_\nu-p_\Delta)^2$, where $p_\nu$ is the incident neutrino momentum and $p_\Delta$ is the LQ momentum.}
$$
y \in \left[\frac{\hat s-m_q^2-m_\Delta^2-\lambda^{1/2}(\sqrt{\hat s},m_\Delta,m_q)}{2 \hat s }
, \frac{\hat s-m_q^2-m_\Delta^2+\lambda^{1/2}(\sqrt{\hat s},m_\Delta,m_q)}{2 \hat s }
\right] \, ,$$
and 
$$x \in \left[ \frac{(m_q+m_\Delta)^2}{2m_NE_\nu},1\right]\, , $$
where $E_\nu$ and $E_\Delta$ are the (anti-)neutrino and
LQ energies in the laboratory frame, $Q^2=-t$ is the
invariant squared momentum transfered in the $t$-channel,
$m_q$ is the mass of the quark in the process, $m_\Delta$ the LQ mass, and
$m_N=(m_p+m_n)/2$ is the average mass of the nucleon. We have used the fact that water is a isoscalar target, so that $N=Z$. We also define the function $\lambda(a,b,c)\equiv(a^2-(b-c)^2)(a^2-(b+c)^2)$.

The two diagrams that contribute to the amplitudes that enter these
cross sections can be seen in Fig.~\ref{fig:gluon}. The cross section
for $\nu_\ell\, g\to q_u\, \Delta^{(-2/3)}$ is given by~\footnote{The
  cross section for the CP conjugated mode $\bar{\nu}_\ell \,g\to
  \bar{q}_u \,\Delta^{(2/3)}$ coincides with this expression.}

\begin{align}
\label{eq:gluon-nu}
\frac{\partial^2\sigma^{g\nu_\ell}}{\partial x \partial y} =(2 m_p E_\nu x) \frac{\alpha_s \vert g_L^{q_u\ell}\vert^2}{8 \hat s}
&\Bigg{\lbrace} \frac{\hat s Q^2+ m_q^2[2(m^2_\Delta+ Q^2)+\hat s] }{(m_q^2+Q^2)^2}\\
&+ \frac{(Q^2+ m^2_\Delta-\hat s)(2 m_\Delta^2+Q^2-\hat s+m_q^2) }{(m_q^2+Q^2-\hat s)^2}
\nonumber \\
&+ \frac{(\hat s-Q^2- m^2_\Delta)(2m_\Delta^2+Q^2)+ m_q^2[\hat s -3(Q^2+m_\Delta^2)] }{(m_q^2+Q^2)(m_q^2+Q^2-\hat s)} \Bigg{\rbrace}\, , \nonumber
\end{align}
where $N_c$ is the number of colors $q_u$ and $\ell$ are a generic up-type quark and lepton, and $\alpha_s=\alpha_s(\hat s)$ is the strong coupling constant at the relevant scale of the process.

\subsection{$\nu_\ell$-quark contribution}

The LQ state contributes to the neutrino-quark cross-section via an exchange of a LQ in the $t$-channel as depicted in Fig.~\ref{fig:nuq}. The SM cross-section ammended with the LQ contribution reads

\begin{align}
\label{eq:cross-section-nuqu}
	\dfrac{\partial^2\sigma^{\nu_\ell {q_u}\to \nu_{\ell^\prime} {q}_{u^\prime}}}{\partial x\partial y} &=  \dfrac{2 m_p E_\nu x}{32 \pi \hat{s}^2}\Bigg{\lbrace} \dfrac{\hat{u}^2\,|g_L^{q_{u}\ell^\prime}|^2 |g_L^{q_{u^\prime}\ell}|^2 }{(\hat{u}-m_\Delta^2)^2}\\[0.4em]
	&+ \delta_{\ell\ell^\prime}\delta_{\hat{u} u^\prime}\Bigg{[}|g_L^{u\ell}|^2\dfrac{16 \sqrt{2} G_f m_Z^2 u^2 (g_A^u-g_V^u)}{(\hat{u}-m_\Delta^2)(\hat{t}-m_Z^2)}\nonumber\\[0.4em]
	&+ \dfrac{8G_F^2m_Z^4 \left[(\hat{s}^2+\hat{u}^2) \left(\left(g_A^u\right)^2+\left(g_V^u\right)^2\right)+2 g_A^u g_V^u (\hat{s}^2-\hat{u}^2)\right]}{(\hat{t}-m_Z^2)^2}\Bigg{]}\Bigg{\rbrace}\,,\nonumber
\end{align}

\noindent where $q_u, q_{u^\prime}$ are generic up-type quarks and $\ell,\ell^\prime$ are generic leptons. We define the weak couplings $g_V^u=T_u^3-2 Q_u \sin^2 \theta_W=1/2-4/3 \sin^2\theta_W$ and $g_A^u=T_u^3=1/2$. The Mandelstam variables in our setup are taken to be
\begin{align}
\hat{s} &= (p_{\nu_\ell}+p_{q_u})^2\,, \qquad \hat{t} = (p_{\nu_\ell}-p_{\nu_{\ell^\prime}})^2\,\quad \text{and} \quad \hat{u} = (p_{\nu_\ell}-p_{q_{u^\prime}})^2.
\end{align}

\noindent Similarly, the $\nu \bar{q}$ cross-section receives a $s$-channel contribution from the LQ state, as shown in Fig.~\ref{fig:nuq}. The corresponding expression is given by

\begin{align}
\label{eq:cross-section-nubarqu}
	\dfrac{\partial^2\sigma^{\nu_\ell \bar{q}_u\to \nu_{\ell^\prime} \bar{q}_{u^\prime}}}{\partial x\partial y}&=  \dfrac{2m_p E_\nu x}{32 \pi \hat{s}^2}\Bigg{\lbrace} \dfrac{\hat{u}^2\,|g_L^{q_{u}\ell^\prime}|^2 |g_L^{q_{u^\prime}\ell}|^2 }{(\hat{s}-m_\Delta^2 )^2+m_\Delta^2\Gamma_{\Delta}^2}\\[0.4em]
	&+ \delta_{\ell\ell^\prime}\delta_{u u^\prime}\Bigg{[} |g_L^{u\ell}|^2\dfrac{16 \sqrt{2} G_f m_Z^2 \hat{u}^2 (g_A^u-g_V^u)(\hat{s}-m_\Delta^2)}{((\hat{s}-m_\Delta^2)^2 +m_\Delta^2\Gamma_{\Delta}^2)(\hat{t}-m_Z^2)}\nonumber\\[0.4em]
	&+ \dfrac{8G_F^2m_Z^4 \left[(\hat{s}^2+\hat{u}^2) \left(\left(g_A^u\right)^2+\left(g_V^u\right)^2\right)+2 g_A^u g_V^u (\hat{s}^2-\hat{u}^2)\right]}{(\hat{t}-m_Z^2)^2}\Bigg{]}\Bigg{\rbrace}\,.\nonumber
\end{align}

\noindent where $\Gamma_{\Delta}$ is the total LQ decay width.

\section{Transport Equations Including the Leptoquark Constribution}
\label{transport}
The set of transport equations that have to be solved in order to compute the attenuation factor for the neutrino flux 
is the following

\bea
\frac{\partial {\cal F}_{\nu_\ell}}{\partial X} &=& -N_A [\sigma_{\nu_\ell N}^{\rm CC}(E) +\sigma_{\nu_\ell N}^{\rm NC}(E)+ 
\sigma_{\nu_\ell}^{\Delta}(E)] {\cal F}_{\nu_\ell}(E,X) \nonumber \\
&+& \frac{1}{c \rho(X)} \int_{0}^{1} \frac{1}{1-y} \left [\frac{d\Gamma_{\tau \to \nu_\ell X'}}{dy}(E_y,y){\cal F}_{\tau}(E_y,X) 
+ \frac{d\Gamma_{\Delta \to \nu_\ell X'}}{dy}(E_y,y){\cal F}_{\Delta}(E_y,X) 
\right]\nonumber \\
&+& N_A \int_{0}^{1} \frac{dy}{1-y} \left [\frac{d\sigma_{\nu_\ell N}^{\rm NC}(E_y,y)}{dy}{\cal F}_{\nu_\ell}(E_y,X) + 
	                                   \frac{d\sigma_{\nu_m \rightarrow \nu_\ell}^{\Delta-q}(E_y,y)}{dy}{\cal F}_{\nu_m}(E_y,X)\right]  \, , \\[0.4em]
\frac{\partial {\cal F}_{\tau}}{\partial X} &=& - \frac{m_\tau}{E\tau_\tau c\rho(X)}{\cal F}_\tau(E,X)+
N_A \int_{0}^{1} \frac{dy}{1-y} \frac{d\sigma_{\nu_\tau N}^{\rm CC}(E_y,y)}{dy} {\cal F}_{\nu_\tau}(E_y,X)\, ,\\[0.4em]
\frac{\partial {\cal F}_{\Delta}}{\partial X} &=& - \frac{m_\Delta}{E\tau_\Delta c\rho(X)}{\cal F}_\Delta(E,X)+
N_A \int_{0}^{1} \frac{dy}{1-y} \frac{d\sigma_{\nu_\beta N}^{\Delta-g}(E_y,y)}{dy} {\cal F}_{\nu_\beta}(E_y,X)\, ,
\eea
where $E_y=E/(1-y)$, $m_\tau$ ($m_\Delta$) is the $\tau$ ($\Delta$) mass, $\tau_\tau$ ($\tau_\Delta$) is the $\tau$ 
($\Delta$) lifetime, $\sigma_{\nu_\ell}^{\Delta}$ is the total LQ contribution to the neutrino-nucleon 
cross section,  $d\sigma_{\nu_\ell}^{\Delta-q}/dy$  ($d\sigma_{\nu_\ell}^{\Delta-g}/dy$)
is the neutrino-quark (neutrino-gluon) part of  the differential cross section 
and $\sigma_{\nu_\ell N}^{\rm CC,NC}$ are the SM CC and NC neutrino-nucleon cross sections.
The distribution of the $\tau$ partial decay widths $d\Gamma_{\tau \to \nu_\ell X'}/dy$ can be found in Ref.~\cite{Dutta:2000jv,Rakshit:2006yi}
while the distribution of the $\Delta$ partial decay width $d\Gamma_{\Delta \to \nu_\ell X'}/dy$ is
given by
\bea
\frac{d\Gamma_{\Delta \to \nu_\ell X'}}{dy}= \Gamma_{\Delta} \frac{(1-y)}{\sqrt{\gamma^2-1}}\gamma \left[ 
\Theta\Big{[}\Big{(}\frac{1}{2}+\frac{\sqrt{\gamma^2-1}}{2\gamma}\Big{)}-y\Big{]}-\Theta\Big{[}\Big{(}\frac{1}{2}-\frac{\sqrt{\gamma^2-1}}{2\gamma}\Big{)}-y\Big{]}
\right] \, ,
\eea

\noindent where $\gamma=E_\Delta/m_\Delta$, $\Theta(x)$ is the Heavyside function.

\section{$B\to D\ell\bar{\nu}$ expressions}
\label{app:flavor}

The contribution of the effective Hamiltonian \ref{eq:hamiltonian} to $\mathcal{B}(B\to D \ell\nu)$ is given by
\begin{align}
\begin{split}
	\label{eq:semilep}
	\dfrac{\mathrm{d}\Gamma}{\mathrm{d}q^2}(B\to D\ell\nu_{\ell}) = \frac{G_F^2|V_{cb}|^2 }{192\pi^3 m_B^3}&|f_+(q^2)|^2 \Bigg{\lbrace} c_+^\ell (q^2)+|g_T(\mu)|^2 c_T^\ell(q^2) \left\vert \frac{f_T(q^2)}{f_+(q^2)}\right\vert^2 \\
	&+c_{TV}^\ell(q^2)\, \mathrm{Re}\left[g_T^\ast(\mu)\right] \frac{f_T(q^2,\mu)}{f_+(q^2)} \\ 
	&+c_0^\ell(q^2) \left\vert 1+g_S(\mu) \frac{q^2}{m_\ell (m_b(\mu)-m_c(\mu))} \right\vert^2 \left\vert\frac{f_0(q^2)}{f_+(q^2)}\right\vert^2 \Bigg{\rbrace},
\end{split}
\end{align}
where $f_{0,+,T}(q^2)$ are the $B\to D$ form factors as defined in Ref.~\cite{Becirevic:2016oho}. The expressions for the phase-space functions $c_i^\ell(q^2)$ read~\cite{Becirevic:2012jf,Becirevic:2016oho}
\begin{align}
	c_+^\ell(q^2) &= \lambda^{3/2}(\sqrt{q^2},m_B,m_D)\left[ 1-\frac{3}{2}\frac{m_\ell^2}{q^2}+\frac{1}{2}\left(\frac{m_\ell^2}{q^2}\right)^3\right], \\
	c_0^\ell (q^2) &= m_\ell^2 \,\lambda^{1/2}(\sqrt{q^2},m_B,m_D)\dfrac{3}{2}\dfrac{m_B^4}{q^2} \left( 1-\dfrac{m_\ell^2}{q^2}\right)^2 \left( 1-\dfrac{m_D^2}{m_B^2}\right)^2, \\
	c_T^\ell(q^2) &=\lambda^{3/2}(\sqrt{q^2},m_B,m_D)\frac{2 q^2}{(m_B+m_D)^2}\left[1-3\left(\frac{m_\ell^2}{q^2}\right)^2+2\left(\frac{m_\ell^2}{q^2}\right)^3\right],\\
	c_{TV}^\ell(q^2)&=\frac{6 m_\ell}{m_B+m_D}\lambda^{3/2}(\sqrt{q^2},m_B,m_D)\left(1-\frac{m_\ell^2}{q^2} \right)^2\, .
\end{align}

\noindent The expressions for $\mathcal{B}(B\to D^\ast \ell \bar{\nu})$ can be found in Ref.~\cite{Sakaki:2013bfa}, which we have independently checked.

\bibliographystyle{JHEP}


\end{document}